\documentclass[
  aps,prd,twocolumn,nofootinbib,superscriptaddress,
  longbibliography,floatfix
]{revtex4-2}

\usepackage{amsmath,amssymb}
\usepackage{bm}
\usepackage{graphicx}
\usepackage{hyperref}
\usepackage{orcidlink}
\definecolor{linknavy}{rgb}{0.0,0.2,0.55}
\hypersetup{
  colorlinks=true,
  linkcolor=linknavy,   
  citecolor=linknavy,   
  urlcolor=linknavy,    
  filecolor=linknavy,
}
\usepackage{tikz}
\usetikzlibrary{feynmangraphz}
\usetikzlibrary{decorations.pathmorphing,decorations.markings,arrows.meta,calc,patterns}
\tikzset{
  midar/.style={postaction={decorate,decoration={markings,
    mark=at position 0.58 with {\arrow{Stealth[length=12.5pt,width=9pt]}}}}},
  nucl/.style={draw,thick,midar},
  helion/.style={draw,thick,double,double distance=2pt,midar},
  phot/.style={draw,thick,
    decorate,decoration={snake,amplitude=2.2pt,segment length=5pt}},
  ar/.style={postaction={decorate,decoration={markings,
    mark=at position #1 with {\arrow{Stealth[length=12.5pt,width=9pt,bend]}}}}},
}
\newcommand{\dwave}[2]{%
  \draw[thick,double,double distance=1.1pt,decorate,
        decoration={snake,amplitude=1.4pt,segment length=6pt}] (#1)--(#2);%
  \draw[draw=none,midar] (#1)--(#2);%
}

\newcommand{\He}{{}^{3}\mathrm{He}}
\newcommand{\Hb}{\overline{{}^{3}\mathrm{He}}}
\newcommand{\dpHe}{d(p,\gamma){}^{3}\mathrm{He}}
\newcommand{\mh}{m_{3}}
\newcommand{\prel}{p_{\mathrm{rel}}}
\newcommand{\Egam}{E_{\gamma}}
\newcommand{\muN}{\mu_{N}}
\newcommand{\ad}[2]{\langle #1 #2\rangle}
\newcommand{\sq}[2]{[#1 #2]}
\newcommand{\Mone}{M1}
\newcommand{\Eone}{E1}
\newcommand{\Etwo}{E2}
\newcommand{\eeff}{e_{\mathrm{eff}}}
\newcommand{\Mbar}{\overline{|\mathcal{M}|^{2}}}
\newcommand{\CSv}{\tilde{C}_{S}}
\newcommand{\CDv}{\tilde{C}_{D}}
\newcommand{\pstar}{p_{\ast}}
\newcommand{\Cone}{\tilde{C}_{1}}
\newcommand{\Ctwo}{\tilde{C}_{2}}
\newcommand{\tEone}{t_{\Eone}}
\newcommand{\tEtwo}{t_{\Etwo}}

\begin{document}

\title{Deuterium-Proton Fusion in an Effective Field Theory\\ Constructed from On-Shell Amplitudes}

\author{Tim M.P.~Tait\,\orcidlink{0000-0003-3002-6909}}
\email{ttait@uci.edu}
\affiliation{Department of Physics and Astronomy,
University of California, Irvine, CA 92697, USA}

\date{\today}

\begin{abstract}
Big Bang nucleosynthesis (BBN) predicts the primordial deuterium abundance to a
precision now limited by the nuclear reactions that burn deuterium. For the
simplest of them, proton--deuteron radiative capture, $d+p \rightarrow \gamma + \He$ [$\dpHe$], the precise
LUNA data sit below the {\it ab initio} benchmark, and BBN reaction networks
split on which to adopt. We develop an effective field theory (EFT) expanding in the finite size of the
nuclei, building the
amplitude with modern on-shell methods that enumerate every tree-level structure consistent with
symmetries without the need for an explicit Lagrangian.  A global Bayesian fit to the capture data and nuclear-theory priors returns
$S(0)=0.209\pm0.008\,\mathrm{eV\,b}$ and traces the offset from the {\it ab initio} benchmark to a
single natural-sized next-to-leading contact term ($\tEone\simeq-0.15$, the
fractional shift of the electric-dipole amplitude)---equivalently a
$\sim\!15\%$ lower effective ${}^{3}\mathrm{He}$ asymptotic normalization.
We estimate the leading EFT truncation errors and identify an elastic $d$--$p$
observable that would separate them. Our results suggest that amplitude methods enable systematic and complete tree-level
construction and matching of EFTs for low-energy nuclear reactions.
\end{abstract}

\maketitle

\section{Introduction}
\label{sec:intro}

The light-element abundances forged in the first minutes after the Big
Bang~\cite{Alpher:1948ve,Wagoner:1966pv} encode the conditions of the early
universe and remain one of the most precise quantitative tests of
cosmology~\cite{Cyburt:2015mya,Pitrou:2018cgg}. Among them,
the primordial deuterium-to-hydrogen ratio $\mathrm{D}/\mathrm{H}$ plays a
special role. It is measured in metal-poor absorption systems to the
percent level, $10^{5}\,\mathrm{D}/\mathrm{H}=2.527\pm0.030$~\cite{Cooke:2017cwo}. And it depends sensitively on the
baryon density $\omega_{b}\equiv\Omega_{b}h^{2}$, providing a determination of
$\omega_{b}$ independent of the cosmic microwave background~\cite{Planck:2018vyg}. The
comparison of the Big Bang nucleosynthesis (BBN) prediction with the measured abundance is a stringent
consistency test of $\Lambda$CDM and a probe of any new physics that
modifies the expansion rate or particle content during
nucleosynthesis~\cite{Pospelov:2010hj}.

The bottleneck in this comparison is no longer observational, but nuclear.
The deuterium abundance is fixed by the competition between deuterium
production and the reactions that burn it, with the predicted $\mathrm{D}/
\mathrm{H}$ most sensitive to the three deuterium-burning rates:
$\dpHe$, $d(d,n){}^{3}\mathrm{He}$, and
$d(d,p){}^{3}\mathrm{H}$~\cite{Cyburt:2015mya,Yeh:2020mgl}. Each is summarized by its
astrophysical $S$-factor,
\begin{equation}
  S(E) = \sigma(E)\,E\,e^{2\pi\eta},
  \qquad \eta \equiv \frac{Z_{1}Z_{2}\,\alpha}{v},
  \label{eq:Sdef}
\end{equation}
which removes the steep Coulomb (Gamow) suppression and the trivial $1/E$
flux factor from the cross section $\sigma(E)$, leaving a smooth function of
the center-of-mass energy $E$ governed by the nuclear matrix element. Thermal
averages of $S(E)$ over the relevant temperatures give the reaction rates
that enter the BBN network~\cite{Wagoner:1966pv,Cyburt:2015mya}. The
uncertainty on these $S$-factors, propagated through the network, dominates
the theoretical error on $\mathrm{D}/\mathrm{H}$~\cite{Mossa:2020gjc}.

Proton--deuteron radiative capture, $\dpHe$, is the only radiative capture
among the three and the cleanest theoretical target. The
reaction is of stellar interest as well: it is the second step of the
proton--proton chain that powers the Sun, burning the deuterium made by $pp$ fusion into ${}^{3}\mathrm{He}$, and its low-energy cross section is a
standard input to solar models and to the modeling of deuterium burning in
pre-main-sequence stars~\cite{Adelberger:2010qa,Acharya:2024lke}. Its cross section has
been measured from the solar Gamow peak up through the BBN window by the LUNA
collaboration, operating deep underground at the Gran Sasso laboratory. There
the rock overburden suppresses the cosmic-ray background by orders of
magnitude, bringing the sub-nanobarn cross sections of the Gamow window into
reach. Using windowless deuterium gas targets, LUNA reached the solar Gamow
peak~\cite{Casella:2002yej} and, at its $400\,\mathrm{kV}$ accelerator,
covered most of the BBN window ($32$--$263\,\mathrm{keV}$) with $2.6\%$
accuracy~\cite{Mossa:2020gjc}. The program is ongoing, most recently inferring
the photon angular distribution across the same
energies~\cite{Stockel:2024hde}. Higher energies have been measured
by several groups~\cite{Tisma:2019acf,Turkat:2021qmq,Schmid:1997zz}.
The benchmark theoretical prediction is the {\it ab initio}
hyperspherical-harmonics calculation of Marcucci {\it et
al.}~\cite{Marcucci:2015yla}, converged at the percent level across the BBN
range. The post-LUNA data nevertheless sit systematically below that
curve~\cite{Pitrou:2020etk,Pisanti:2020efz}, and the reaction networks split on
how to respond. PRIMAT adopts a scaled version of the {\it ab initio}
curve~\cite{Pitrou:2018cgg,Moscoso:2021xog}; PArthENoPE uses a polynomial fit to the
data~\cite{Gariazzo:2021iiu}.  The difference propagates into the predicted
$\mathrm{D}/\mathrm{H}$, leading to a $\sim 2\sigma$ discrepancy between the two approaches. 
A data-driven Gaussian-process analysis recently
determined the energy interval over which the rate must be known: outside
$E\in[14,600]\,\mathrm{keV}$, truncating the rate integral shifts
$\mathrm{D}/\mathrm{H}$ by less than $0.1\sigma$ of the observational
uncertainty~\cite{Launders:2026ciu}. The BBN Gamow peak sits near
$100\,\mathrm{keV}$.

In this paper we approach $\dpHe$ via a nuclear-state effective field theory built with on-shell amplitude
methods. 
The nuclei are treated 
as point-like degrees of freedom, their finite size restored order by order through short-range contact interactions
organized into a controlled low-energy expansion.
At BBN energies the relative momentum $\prel$ lies well below the
deuteron-breakup scale $\pstar\approx53\,\mathrm{MeV}$, the nearest omitted
dynamics. This justifies integrating out the substructure. What remains is
a contact EFT of point-like nuclear states whose interactions are short-range operators ordered by
powers of $\prel/\pstar$. In this organizing principle it is similar in spirit to the
halo and cluster EFTs~\cite{Bertulani:2002sz,Bedaque:2003wa,Rupak:2016mmz}. It is
not itself a cluster EFT, however: rather than resolving ${}^{3}\mathrm{He}$ into a $p+d$ cluster, we
construct the capture amplitude directly from on-shell data. Massive spinor-helicity
variables, on-shell recursion, and leading boundary terms generate every tree-level structure
consistent with the symmetries, without reference to a particular Lagrangian.

This calculation is, to our knowledge, the first time the modern on-shell
program has been carried through to a nuclear astrophysics observable with
quantified uncertainties.  
We choose $\dpHe$ because it is anchored
by both precise experimental data and sophisticated {\it ab initio} nuclear predictions, allowing
a controlled assessment of the EFT/amplitude methods in the context of nuclear reactions.
More broadly, {\it ab initio} reaction theory built on chiral effective field
theory~\cite{Epelbaum:2008ga,Machleidt:2011zz} has recently reached a range of neighboring
light-nucleus radiative captures of primordial and stellar interest---among them
$d(\alpha,\gamma){}^{6}\mathrm{Li}$~\cite{Hebborn:2022iiz} and
${}^{3}\mathrm{He}(\alpha,\gamma){}^{7}\mathrm{Be}$~\cite{Atkinson:2024zrm}---sharpening the
benchmarks available to methods such as this one. A reaction-level chiral-EFT treatment of the
$\dpHe$ $S$-factor itself is still lacking: chiral-interaction calculations of nucleon--deuteron
radiative capture~\cite{Skibinski:2006gy} address higher energies and polarization observables
rather than the threshold astrophysical regime, and the benchmark we compare
to~\cite{Marcucci:2015yla} is based on conventional phenomenological interactions.

We find that the on-shell EFT construction succeeds: the pipeline runs
from construction of the on-shell vertices to a gauge-invariant $S$-factor, the multipole tower is reproduced, 
and the map between the $S$-factor coefficients and the underlying
low energy constants is derived. The mapping extracts the combinations of nuclear
inputs that the $\dpHe$ capture data determine. The theory-versus-data offset emerges
as a single low-energy coefficient of natural EFT size, degenerate with the
${}^{3}\mathrm{He}$ asymptotic normalization. The construction also identifies an additional observable
whose future measurement would help separate the two.
The physics deliverables are thus a data-anchored $S(E)$ with quantified
truncation uncertainty and a testable diagnosis of the {\it ab initio}--data
offset, of direct relevance to how the BBN network codes treat this reaction
(Sec.~\ref{sec:outlook}). The methodological deliverable is the demonstration
that on-shell amplitude methods produce a complete, systematically improvable
tree-level EFT for a low-energy nuclear reaction.
Of course, an EFT can never make predictions beyond the resolving power of the data or theory
inputs used to match it to observations.  Its value lies in identifying structural features and relations between observables, providing a systematic framework for improving precision, and estimating uncertainties from the undetermined higher-order coefficients.

We intend this paper as the first installment in a series. The present
reaction is where the methodology can be validated step by step. Only then
will it be pointed at the transfer channels $d(d,n){}^{3}\mathrm{He}$ and $d(d,p){}^{3}\mathrm{H}$,
which are less precisely measured and lack an {\it ab initio} benchmark of comparable maturity.

The remainder of the paper is organized as follows.
Section~\ref{sec:eft} develops the EFT in the context of the on-shell amplitude
program. Section~\ref{sec:dpHe} presents the $\dpHe$ $S$-factor, its
multipole content, and its matching to measured nuclear inputs.
Section~\ref{sec:results} determines the low-energy constants by a global fit to the
data, inferring the best Bayesian determination of $S(E)$ in light of the experimental and theoretical inputs.  
Section~\ref{sec:outlook} contains discussion and
outlook.  Details of the amplitude construction are deferred to Appendix~\ref{app:amp}, and an alternate completely
data-driven fit is presented in Appendix~\ref{app:efffit}.

\section{Effective Field Theory and the On-Shell Amplitude Program}
\label{sec:eft}

\subsection{EFT Approach}
\label{sec:why}

Effective field theory is the systematic way to describe physics at a low scale $p$ when the relevant ultraviolet dynamics lives at a higher scale
$\Lambda$. The description is an expansion in powers of $p/\Lambda$, with a
finite number of low-energy constants (LECs) fixed at each order from data or
a matching calculation~\cite{Hammer:2019poc}. For two-nucleon and few-nucleon
systems this program has a mature realization in pionless and halo/cluster
EFTs~\cite{Kaplan:1998we,vanKolck:1998bw,Chen:1999tn,Hammer:2019poc}, where the expansion parameter is
the ratio of the typical momentum to the breakdown scale set by the first
omitted physics.
In the present work, we construct an EFT with a separation of length scales
in which the internal structure of the nuclei is encoded as a short-distance
effect, and the degrees of freedom are the relevant nuclear states,
characterized at long distances by their masses, spins, charges, and moments.
Nuclear interactions rearranging the nuclei and the short-range electroweak
interactions --- moments, two-body currents, finite-size effects --- are
represented by local interactions among these degrees of freedom.

Radiative capture at BBN energies is a favorable setting for such an expansion. 
The kinematics of $\dpHe$ are characterized by the ${}^{3}\mathrm{He}$ proton separation energy,
\begin{equation}
  Q = m_{p}+m_{d}-\mh = 5.4934\,\mathrm{MeV},
\end{equation}
with the emitted photon carrying $\Egam\simeq Q+E\approx5.5$--$6.1\,\mathrm{MeV}$ across the window (see Table~\ref{tab:inputs}).
The entrance reduced mass is $\mu=m_{p}m_{d}/(m_{p}+m_{d})\approx
625\,\mathrm{MeV}$, giving a relative momentum $\prel=\sqrt{2\mu E}$ of
$4$--$27\,\mathrm{MeV}$ for $E=14$--$600\,\mathrm{keV}$.

Rather than resolving ${}^{3}\mathrm{He}$ into three nucleons bound by a
potential, we take the deuteron, proton, and ${}^{3}\mathrm{He}$ states themselves
as the degrees of freedom. A strong $d$--$p$--${}^{3}\mathrm{He}$
coupling then describes the compositeness, including the asymptotic normalization
coefficient (ANC) of ${}^{3}\mathrm{He}$ in the $p+d$ channel.
This is similar to the logic of cluster EFT, applied to other
radiative-capture reactions such as the recently studied
$d(\alpha,\gamma){}^{6}\mathrm{Li}$~\cite{Nazari:2024rpi}.
Pionless EFT has been applied to neighboring electromagnetic and weak captures of
primordial and solar interest, among them $np\to d\gamma$~\cite{Chen:1999bg} and $pp$
fusion~\cite{Kong:2000px}. The nearest existing EFT analysis of $\dpHe$ itself works at
the nucleon level, targeting the solar Gamow peak~\cite{Sadeghi:2008zz}.

The effective theory is organized by a single physical scale, fixed by the nearest singularity of
the capture amplitude---not the pion, but the deuteron breakup. The entrance channel couples
virtually, through deuteron distortion, to the $p+p+n$ continuum at
$E=+B_{d}=2.2246\,\mathrm{MeV}$. A Taylor expansion of $S(E)$ about
threshold therefore converges only for $|E|<B_{d}$, i.e.\ in the variable $E/B_{d}=(\prel/\pstar)^{2}$ with
$\pstar=\sqrt{2\mu B_{d}}\approx53\,\mathrm{MeV}$. Across the window this
variable runs from $0.045$ at the Gamow peak to $0.27$ at the top. The same breakup physics sizes the short-range operators. A
contact term encodes the breakup loop, so it carries a loop factor relative to a tree-level
operator at $\pstar$. Its natural magnitude corresponds to an effective cutoff of order
$\Lambda\sim\sqrt{4\pi}\,\pstar\approx187\,\mathrm{MeV}$. We adopt this
$\Lambda$ as the explicit normalization scale of the contact couplings below,
which are then dimensionless. That $\Lambda$ is comparable to the
pion mass $m_{\pi}\approx140\,\mathrm{MeV}$, the breakdown scale of pionless
EFT, is the familiar statement that the deuteron is a
natural shallow bound state. The estimate, however, needs only the breakup scale $\pstar$ and a
generic loop factor, with no reference to the pion. The single scale $\pstar$ thus sets the radius
of convergence directly, and $\Lambda\sim\sqrt{4\pi}\,\pstar$ fixes the size of
the short-range operators.
A second internal scale, the ${}^{3}\mathrm{He}$ binding
momentum $\kappa=\sqrt{2\mu Q}\approx83\,\mathrm{MeV}$, is likewise carried by the
LECs.

We build the amplitude at tree level, with no $d$--$p$ rescattering in the entrance
channel. This omission must be justified channel by channel: the $s$-wave
$d$--$p$ interactions are not weak (the quartet scattering length is
$a_{4}\approx14\,\mathrm{fm}$~\cite{Rupak:2001ci,Kievsky:1997zz}, and the doublet has a delicate
near-threshold pole~\cite{Bedaque:1998kg,Kievsky:1997zz}). A full pionless-EFT treatment resums these
interactions and describes the $pd$/${}^{3}\mathrm{He}$ system and related $A=3$ observables
directly~\cite{Vanasse:2014kxa,Konig:2015aka,Vanasse:2016jtc,Vanasse:2015fph}.
The protection is a separation of regimes. The electric dipole proceeds through
$p$-wave entrance phases, which are small across the window, so its rescattering is a
higher-order correction absorbed into $c_{\Eone}$~\cite{Rupak:2016mmz}. The magnetic dipole is sensitive to
the large $s$-wave lengths, but it is appreciable only near threshold, where $\prel a$ is
itself small. Its short-range strength is in any case fixed from data
(Sec.~\ref{sec:results}) rather than predicted.

\subsection{Amplitudes for EFTs}
\label{sec:onshell}

Building an EFT amplitude by writing a Lagrangian and gauge-fixing it is
workable but redundant: many operators related by field redefinitions and
equations of motion describe the same physics, and for higher-spin fields
such as the deuteron the off-shell vertices proliferate. The on-shell
program sidesteps this by constructing amplitudes directly from their
physical content. In the massive spinor-helicity formalism of
Arkani-Hamed, Huang, and Huang~\cite{Arkani-Hamed:2017jhn}, every external
leg of spin $S$ carries a symmetric rank-$2S$ little-group tensor (indicated as bold
spinors). Any local three-point amplitude is then a polynomial in the
pairwise spinor brackets, its form fixed up to a finite number of
coupling constants by Lorentz invariance and the little-group weights of
the legs~\cite{Arkani-Hamed:2017jhn,Cheung:2017pzi}. Massive higher-spin couplings, which are
awkward in the Lagrangian language, are then simply the independent bracket
structures of the correct weight.

Two features make this construction the right tool for the present problem.
First, it enumerates {\it all} tree-level structures consistent with the
symmetries at a given order, so the EFT operator basis is complete by
construction. The strong vertex yields exactly the two parity-even
$d$--$p$--${}^{3}\mathrm{He}$ couplings, an $s$-wave and a $d$-wave. The
photon couplings of the spin-$\tfrac12$ and spin-$1$ nuclei reduce to the
charge, magnetic-moment, and (for the deuteron) quadrupole form factors,
with the charge fixed by the Ward identity. Second, four-point amplitudes
are assembled from the three-point data by on-shell recursion: the
Britto--Cachazo--Feng--Witten relations~\cite{Britto:2004ap,Britto:2005fq}
reconstruct the factorizable part of the amplitude from its residues on
physical poles, where it factorizes into products of three-point
amplitudes. For a radiative process the recursion is supplemented by a
contact (boundary) term that is not visible on any factorization channel.
This term carries the electromagnetic two-body currents, fixed in part
by the Ward identity and in part by a finite set of free LECs at a given order in the EFT expansion. The complete
enumeration of these boundary structures at a given order is an EFT operator-counting
problem, which we solve using the helicity-category algorithm of Ref.~\cite{DeAngelis:2022qco}.

\section{Proton--Deuteron Radiative Capture}
\label{sec:dpHe}

We construct the $\dpHe$ amplitude with on-shell methods, as a sum of factorized
(pole) channels and local boundary terms (Fig.~\ref{fig:diagrams}). The explicit
construction is given in Appendix~\ref{app:amp}. The factorized channels
[Fig.~\ref{fig:diagrams}(a)--(c)] glue on-shell three-point vertices across the
single-particle pole of each channel, one for the photon radiated from each
charged leg (the outgoing ${}^{3}\mathrm{He}$, the incoming proton, or the
incoming deuteron). Their residues are therefore fixed by the nuclear couplings
and the Ward-determined charges. What the
poles leave undetermined is a local polynomial remainder: the boundary terms
[Fig.~\ref{fig:diagrams}(d)], contact operators encapsulating nuclear substructure, including the electromagnetic two-body
currents. One is fixed by the Ward identity, combining with the factorized pieces to restore gauge invariance. 
The rest are free Wilson coefficients: the leading $\Mone$,
$\Eone$, and $\Etwo$ contact terms. 

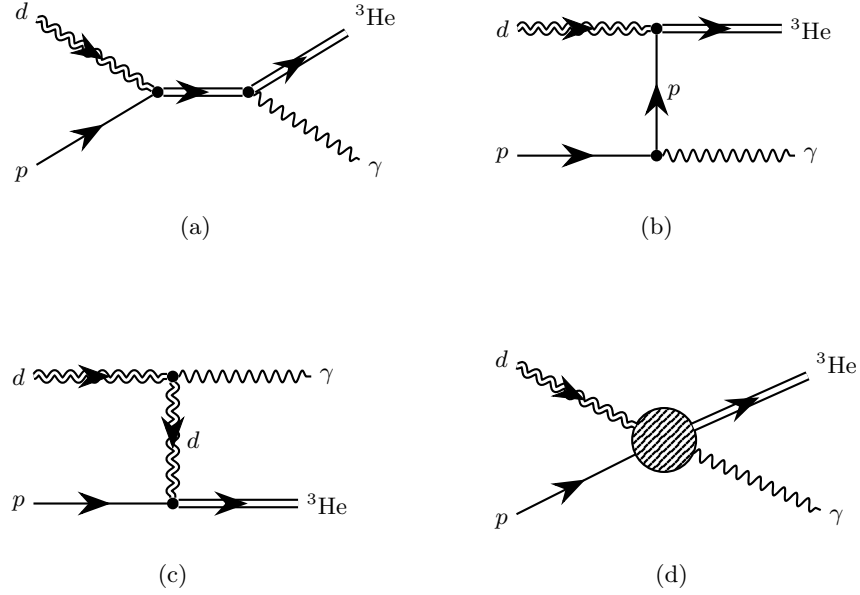
\begin{figure*}[t]
\centering
\begin{tikzpicture}[fmg, every node/.append style={font=\small}]
\begin{scope}[local bounding box=A]
  \path (-0.3,1.08) node (d) {$d$}
        (-0.3,-1.08) node (p) {$p$}
        (1.5,0) node[vertex,name=S] {}
        (2.7,0) node[vertex,name=E] {}
        (4.38,1.02) node (he) {${}^3\mathrm{He}$}
        (4.38,-1.02) node (g) {$\gamma$};
  \dwave{d}{S}
  \draw[nucl]   (p) -- (S);
  \draw[helion] (S) -- (E);
  \draw[helion] (E) -- (he);
  \draw[phot]   (E) -- (g);
  \node at (2.0,-1.8) {(a)};
\end{scope}
\begin{scope}[shift={(6.4,0)},local bounding box=B]
  \path (-0.34,0.84)  node (d) {$d$}
        (-0.34,-0.84) node (p) {$p$}
        (1.7,0.84)  node[vertex,name=S] {}
        (1.7,-0.84) node[vertex,name=E] {}
        (3.74,0.84)  node (he) {${}^3\mathrm{He}$}
        (3.74,-0.84) node (g) {$\gamma$};
  \dwave{d}{S}
  \draw[nucl]   (p) -- (E);
  \draw[nucl]   (E) -- (S);
  \draw[helion] (S) -- (he);
  \draw[phot]   (E) -- (g);
  \node[right=1pt] at (1.7,0) {$p$};
  \node at (1.7,-1.8) {(b)};
\end{scope}
\begin{scope}[shift={(0,-4.6)},local bounding box=C]
  \path (-0.34,0.84)  node (d) {$d$}
        (-0.34,-0.84) node (p) {$p$}
        (1.7,0.84)  node[vertex,name=E] {}
        (1.7,-0.84) node[vertex,name=S] {}
        (3.74,0.84)  node (g) {$\gamma$}
        (3.74,-0.84) node (he) {${}^3\mathrm{He}$};
  \dwave{d}{E}
  \draw[nucl]   (p) -- (S);
  \dwave{E}{S}
  \draw[phot]   (E) -- (g);
  \draw[helion] (S) -- (he);
  \node[right=2pt] at (1.7,0) {$d$};
  \node at (1.7,-1.8) {(c)};
\end{scope}
\begin{scope}[shift={(6.4,-4.6)},local bounding box=D]
  \path (-0.35,1.08) node (d) {$d$}
        (-0.35,-1.08) node (p) {$p$}
        (1.8,0) node[blob,minimum size=2.6em,name=Bl] {}
        (4.08,1.02) node (he) {${}^3\mathrm{He}$}
        (4.08,-1.02) node (g) {$\gamma$};
  \dwave{d}{Bl}
  \draw[nucl]   (p) -- (Bl);
  \draw[helion] (Bl) -- (he);
  \draw[phot]   (Bl) -- (g);
  \node at (1.9,-1.8) {(d)};
\end{scope}
\end{tikzpicture}
\caption{Contributions to $\dpHe$ in the nuclear-state amplitude EFT.
Diagrams (a)--(c) are the factorized channels, with the internal line the
propagator whose pole defines the channel. Diagram (d) is the contact (boundary) term, including the piece required by the Ward identity and higher-dimension operators.
The deuteron is denoted by a double wavy line, ${}^{3}\mathrm{He}$ by a double line, and the proton by a single line,
all with arrows indicating the flow of the nuclei; the single wavy line is
the photon.}
\label{fig:diagrams}
\end{figure*}

\subsection{$S$-Factor: Multipole Analysis}
\label{sec:sfactor}

The capture process $d(p_{d})+p(p_{p})\to{}^{3}\mathrm{He}(p_{3})+
\gamma(p_{\gamma})$ is a $2\to2$ reaction with a single charged pair in the
entrance channel. 
The gauge-invariant matrix element is assembled in
Appendix~\ref{app:amp}. Squaring it, averaging over initial and summing over
final spins, and integrating over the photon emission angle gives the incoherent
multipole sum of Eq.~\eqref{eq:Msq}. Cross-multipole interferences average to
zero. Each photon multipole then
enters through its own squared strength
$\mathcal{A}_{X}=N_{X}\,\Egam^{2L_{X}}|U_{X}|^{2}$
[Eqs.~\eqref{eq:AM1}--\eqref{eq:AE2}], weighted by the centrifugal factor
$(\prel^{2}/\mh^{2})^{L}$ of its entrance partial wave.
$\prel^{2}=2\mu E$ is the relative incoming momentum, and the
photon energy is
$\Egam=|\vec p_{\gamma}|=(s-\mh^{2})/(2\sqrt{s})\simeq Q+E$ with
$s=(m_{d}+m_{p}+E)^{2}$.

After the Sommerfeld removal of Eq.~\eqref{eq:Sdef}, the higher partial waves retain a residual
Coulomb-penetration remnant that the $s$-wave subtraction leaves behind: the
$\Eone$ (and $M2$) $p$-wave normalization carries an extra factor $(1+\eta^{2})$,
$\eta^{2}=\alpha^{2}\mu/2E$, and the $\Etwo$ $d$-wave a factor
$(1+\eta^{2})(1+\eta^{2}/4)$. This remnant is around $+17\%$ at the $100\,\mathrm{keV}$ Gamow peak and grows
toward threshold. The growth is harmless: the physical combination
$E(1+\eta^{2})=E+E_{C}$, with $E_{C}=\alpha^{2}\mu/2=16.7\,\mathrm{keV}$, stays
finite, so the electric dipole does not vanish at threshold.
The resulting $S$-factor takes the form
\begin{equation}
\begin{split}
  S(E) = \frac{\alpha\,\Egam}{48\,s}
  \Big[\,& A_{\Mone} + A_{\Etwo}^{D}
    + \big(A_{\Eone}+A_{M2}\big)\tfrac{2\mu (E+E_{C})}{\mh^{2}} \\
  & + A_{\Etwo}\big(\tfrac{2\mu}{\mh^{2}}\big)^{2}(E+E_{C})(E+\tfrac{E_{C}}{4})
  \Big],
\end{split}
  \label{eq:SE}
\end{equation}
consisting of the magnetic-dipole ($\Mone$, $s$-wave) term, joined at threshold
by the small $D$-state quadrupole $A_{\Etwo}^{D}$, the $p$-wave electric dipole
with its subleading magnetic-quadrupole companion ($\Eone,M2$), and the
$d$-wave electric quadrupole ($\Etwo$).
The coefficients are
\begin{equation}
\begin{aligned}
  A_{\Mone} &= \Egam^{2}\big(\tfrac14|U_{\Mone}^{(2)}|^{2}
              +\tfrac29|U_{\Mone}^{(4)}|^{2}\big),\\
  A_{\Eone} &= \tfrac13\,\Egam^{2}\,|U_{\Eone}|^{2},\\
  A_{M2}    &= \Egam^{4}\big(N_{M2}^{(a)}|U_{M2}^{(a)}|^{2}
              +N_{M2}^{(b)}|U_{M2}^{(b)}|^{2}\big),\\
  A_{\Etwo} &= \Egam^{4}\big(\tfrac13|U_{\Etwo}|^{2}
              +\tfrac{2}{21}|U_{\Etwo}^{(D)}|^{2}\big),\\
  A_{\Etwo}^{D} &= \tfrac{1}{15}\,\Egam^{4}\,|U_{\Etwo}^{(D)}|^{2},
\end{aligned}
  \label{eq:Aexplicit}
\end{equation}
with reduced matrix elements
\begin{equation}
\begin{aligned}
  U_{\Mone}^{(2)} &= \frac{\CSv}{m_{d}}\big(2\mu_{3}P_{3}+\tfrac43\mu_{d}P_{d}
                     -\tfrac23\mu_{p}P_{p}\big)+\frac{e}{\Lambda^{3}}\,c_{\Mone}^{(2)},\\
  U_{\Mone}^{(4)} &= \frac{\CSv}{m_{d}}\big(\mu_{d}P_{d}-2\mu_{p}P_{p}\big)+\frac{e}{\Lambda^{3}}\,c_{\Mone}^{(4)},\\
  U_{\Eone}       &= \frac{\eeff\,\CSv}{m_{d}}+\frac{e}{\Lambda}\,c_{\Eone},\\
  U_{M2}^{(a)}    &= \frac{e}{\Lambda^{3}}\,c_{M2}^{(a)},\qquad U_{M2}^{(b)}=\frac{e}{\Lambda^{3}}\,c_{M2}^{(b)},\\
  U_{\Etwo}       &= \frac{q_{\mathrm{eff}}\,\CSv}{m_{d}}+\frac{e}{\Lambda}\,c_{\Etwo},\\
  U_{\Etwo}^{(D)} &= \frac{q_{\mathrm{eff}}\,\CDv}{m_{d}},
\end{aligned}
  \label{eq:Udefs}
\end{equation}
containing threshold propagators $P_{X}=1/(2m_{X}\Egam)$, magnetic moments $\mu_{X}$,
recoil charges $\eeff,q_{\mathrm{eff}}$, the strong-vertex $s$- and $d$-wave
couplings $\CSv,\CDv$ (the on-shell vertex couplings of Appendix~\ref{app:vertices},
fixed by the ${}^{3}\mathrm{He}$ ANCs $C_{S},C_{D}$ through the relation of
Sec.~\ref{sec:matching}), and short-range contact terms $c_{X}$ detailed in
Sec.~\ref{sec:matching} and Appendix~\ref{app:me}. All couplings are
dimensionless: the spin-$1$ polarization factor $1/m_{d}$ of the deuteron leg
and the contact normalization $e/\Lambda^{n}$, with 
\begin{equation}
\Lambda  =  \sqrt{4\pi}\,\pstar = 187.0~\mathrm{MeV},
\end{equation}
are written explicitly, each contact term
carrying the power $n$ fixed by the dimension of its reduced matrix element,
the order-one constants of the multipole projection being absorbed into the
$c_{X}$.
The $\mu_{X}$ are the dimensionless moments in nuclear magnetons
(Table~\ref{tab:inputs}) and $[P_{X}]=-2$, so the reduced matrix elements carry
their dimensions in the explicit kinematic factors alone
($[U_{\Mone}]=-3$, $[U_{\Eone}]=[U_{\Etwo}]=-1$; all quantities in MeV units).

Each term carries the photon weight $\Egam^{2L_{X}}$: $\Egam^{2}$ for the
dipoles and $\Egam^{4}$ for the quadrupoles. The dominant energy
dependence thus factors into a fixed kinematic prefactor, leaving a
short polynomial in $E$. Writing the energy-independent reduced strengths
$\bar A_{X}\equiv A_{X}/\Egam^{2L_{X}}$ and pulling
out the shared dipole weight $\Egam^{2}$ and the phase-space $\Egam$,
\begin{equation}
\begin{aligned}
  S(E)&=P(E)\,\big[\,a_{0}+a_{1}E+a_{2}E^{2}+\cdots\,\big],\\
  P(E)&=\frac{\alpha\,\Egam^{3}}{48\,s}
       =\frac{\alpha\,(Q+E)^{3}}{48\,(m_{d}+m_{p}+E)^{2}},
\end{aligned}
  \label{eq:poly}
\end{equation}
with the prefactor $P(E)$ carrying the kinematic curvature, and Taylor
coefficients
\begin{equation}
\begin{aligned}
  a_{0}&=\bar A_{\Mone}+\tfrac{2\mu E_{C}}{\mh^{2}}\bar A_{\Eone}
        +Q^{2}\bar A_{\Etwo}^{D}+\cdots,\\
  a_{1}&=\tfrac{2\mu}{\mh^{2}}\big(\bar A_{\Eone}+Q^{2}\bar A_{M2}\big)
        +2Q\,\bar A_{\Etwo}^{D}+\cdots,\\
  a_{2}&=\big(\tfrac{2\mu}{\mh^{2}}\big)^{2}Q^{2}\bar A_{\Etwo}
       +\tfrac{4\mu Q}{\mh^{2}}\,\bar A_{M2}+\bar A_{\Etwo}^{D}+\cdots,
\end{aligned}
  \label{eq:taylor}
\end{equation}
collecting the expansion in $E/Q$.  The $\cdots$ represent smaller $E_{C}$ cross-terms
and higher orders in $E/Q$.
The $E_{C}$-enhanced $\Eone$ tail enters the threshold
coefficient $a_{0}$ alongside the magnetic-dipole strength $|U_{\Mone}|^{2}$,
so $S(0)=P(0)\,a_{0}$ does \emph{not} isolate the magnetic dipole. The slope
$a_{1}$ is the electric-dipole strength proper and $a_{2}$ the electric
quadrupole. The constant $D$-state piece
$\bar A_{\Etwo}^{D}\sim(C_{D}/C_{S})^{2}$ is the smallest threshold term. 

The $a_{i}$ of Eq.~(\ref{eq:taylor}) are the
coefficients of the polynomial multiplying $P(E)$ as opposed to the coefficients of the Taylor expansion
of $S(E)$ itself. The conventional low-energy expansion
$S(E)=S(0)+S'(0)E+\tfrac12 S''(0)E^{2}+\cdots$ follows by expanding $P(E)$ and collecting powers:
\begin{equation}
\begin{aligned}
  S(0)&=P(0)\,a_{0},\qquad S'(0)=P(0)\,a_{1}+P'(0)\,a_{0},\\
  \tfrac12 S''(0)&=P(0)\,a_{2}+P'(0)\,a_{1}+\tfrac12 P''(0)\,a_{0},
\end{aligned}
  \label{eq:Sderiv}
\end{equation}
with $P'(0)/P(0)=3/Q-2/(m_{d}+m_{p})$ and
$P''(0)/P(0)=6\big(1/Q-1/(m_{d}+m_{p})\big)^{2}\simeq6/Q^{2}$, both fixed by the
masses and the $Q$-value.
The curvature $S''(0)$ is thus dominated by the kinematic $P''(0)a_{0}$
term, the exact $\Egam^{3}/s$ rise. The genuine electric-quadrupole
and effective-range content $P(0)a_{2}$ sits at the percent level.

The multipole hierarchy has a clean physical consequence, sharpened by the Coulomb
remnant. The magnetic dipole is unsuppressed at threshold. But so is the electric
dipole: the Coulomb remnant keeps its tail finite at $E=0$, so the two are
comparable there ($\Eone/\Mone\approx1$ with the fitted strengths of
Sec.~\ref{sec:results}). The electric dipole already dominates by the bottom of
the window and grows steeply ($\Eone/\Mone\approx7$ at $100\,\mathrm{keV}$,
$\approx36$ at the top). The steep rise of the $\dpHe$ $S$-factor is thus the
electric dipole switching on. The threshold normalization, by contrast, is \emph{shared}
roughly equally between the magnetic dipole and the Coulomb-enhanced
electric-dipole tail. This comparable-at-threshold split is the established
phenomenology: the near-threshold angular distributions of Schmid
{\it et al.}~\cite{Schmid:1997zz}, the LUNA angular-distribution measurement
across the BBN window~\cite{Stockel:2024hde}, and the {\it ab initio} multipole
decomposition all place a substantial ($\sim$half) electric-dipole share in
$S(0)$. It also serves as a check on the EFT counting. The corollary for the
analysis: solar/threshold data fix the \emph{sum} of the magnetic strength and the
electric-dipole tail (the correlated $a_{0}$--$a_{1}$ combination), not the
magnetic dipole alone. The BBN-window data then pin the electric-dipole slope.

Section~\ref{sec:results} determines the low-energy constants by a global fit that combines the capture data with the theory inputs (the ${}^{3}\mathrm{He}$ ANC and naive dimensional analysis (NDA)~\cite{Manohar:1983md} priors), with the leading multipoles fixed by the construction (Sec.~\ref{sec:matching}). The same functional form also supports a purely data-driven analysis, a three-coefficient fit of $\{a_{0},a_{1},a_{2}\}$ to the data using none of the nuclear-structure input, which delivers $S(E)$ and its covariance directly; we present that fit in Appendix~\ref{app:efffit}.

\subsection{Matching}
\label{sec:matching}

The value of the on-shell construction is that each multipole strength in
Eq.~\eqref{eq:SE} reduces to a reduced matrix element built out of the
LECs, so the $S$-factor coefficients become explicit functions of measured
electromagnetic moments, the strong $d$--$p$--${}^{3}\mathrm{He}$ couplings,
and short-range contact terms. The external inputs and their
provenance are collected in Table~\ref{tab:inputs}.

\begin{table}[t]
\caption{External inputs to the $\dpHe$ $S$-factor. Masses are from
CODATA~\cite{Tiesinga:2024CODATA} and AME2020~\cite{Wang:2021xhn}, magnetic
moments from PDG~\cite{ParticleDataGroup:2024cfk}
and CODATA~\cite{Tiesinga:2024CODATA}, the deuteron quadrupole from the
molecular electric-field-gradient determination of
Ref.~\cite{Pavanello:2010mzw}, and the deuteron asymptotic $D/S$ ratio
$\eta_{D}$ from measurement~\cite{Rodning:1990gg}.}
\label{tab:inputs}
\begin{ruledtabular}
\begin{tabular}{l c l}
Quantity & Value & Role / source \\
\hline
$m_{p}$            & $938.272\,089\,\mathrm{MeV}$ & CODATA \\
$m_{d}$            & $1875.612\,943\,\mathrm{MeV}$ & CODATA \\
$\mh$              & $2808.391\,607\,\mathrm{MeV}$ & AME2020 \\
$\mu_{p}$          & $+2.792\,847\,\muN$ & PDG + CODATA\\
$\mu_{d}$          & $+0.857\,438\,\muN$ & PDG + CODATA\\
$\mu_{3}$        & $-2.127\,625\,\muN$ & PDG + CODATA\\
$\mathcal{Q}_{d}$  & $0.285\,783\,e\,\mathrm{fm}^{2}$ & Ref.~\cite{Pavanello:2010mzw} \\
$\eta_{D}$         & $0.0256(4)$ & ~Ref.~\cite{Rodning:1990gg} \\
$C_{S}$            & $2.144\,\mathrm{fm}^{-1/2}$ & ${}^{3}\mathrm{He}$ $p$+$d$ $s$-wave ANC~\cite{Nollett:2011qf} \\
$C_{D}/C_{S}$      & $-0.0432$ & ${}^{3}\mathrm{He}$ $p$+$d$ $D/S$ ratio~\cite{Nollett:2011qf,Kievsky:1997zz} \\
\end{tabular}
\end{ruledtabular}
\end{table}

\subsubsection{Strong Couplings: $C_{S}$ and $C_{D}$}

The strong $d$--$p$--${}^{3}\mathrm{He}$ couplings encode important
nuclear-structure input that the EFT cannot predict. 
Both are related to asymptotic normalizations of the
${}^{3}\mathrm{He}$ ground state in the $p+d$ channel: $C_{S}$ the $s$-wave,
$C_{D}$ the $d$-wave. They are fixed from {\it ab initio} few-body
theory~\cite{Nollett:2011qf,Kievsky:1997zz} and the measured $D/S$ ratio.

$C_{S}$ is the dimensionful $s$-wave ANC, set by the Whittaker tail of the
$p$--$d$ overlap, $R(r)\to C_{S}\,W_{-\eta_{b},1/2}(2\kappa r)/r$, with binding
momentum $\kappa=\sqrt{2\mu Q}=0.420\,\mathrm{fm}^{-1}$ ($82.9\,\mathrm{MeV}$;
the internal scale of Sec.~\ref{sec:eft})
and bound-state Coulomb parameter $\eta_{b}=0.055$. The variational Monte Carlo determination of Nollett and Wiringa,
$C_{S}=2.144(8)\,\mathrm{fm}^{-1/2}$~\cite{Nollett:2011qf}, and the Pisa
hyperspherical-harmonics solution~\cite{Kievsky:1997zz} agree to a couple of
percent.\footnote{Ref.~\cite{Nollett:2011qf} does not extract the ANC from the explicit
overlap integral, whose long-range tail is only weakly determined by the variational
Monte Carlo wavefunction (optimized for the energy). It instead uses a
Green's-function integral relation weighted toward the nuclear interior and insensitive
to that tail, evaluated at the physical separation energy even when the
Hamiltonian's binding differs slightly. Ref.~\cite{Friar:1988zz} defines the deuteron-normalized
Friar--Gibson--Payne {\it channel} constant, whose AV18+UR value is
$1.88$~\cite{Kievsky:1997zz}. The {\it overlap}
ANC entering the amplitude follows by the unit factor $\sqrt{2\kappa}=0.92\,\mathrm{fm}^{-1/2}$
and the $\sqrt{3/2}$ spectroscopic factor of ${}^{3}\mathrm{He}\to d+p$, giving
$1.88\to2.1\,\mathrm{fm}^{-1/2}$.}

The reduced matrix elements of Eq.~\eqref{eq:Udefs} are expressed not in terms of the ANC directly,
but in terms of the dimensionless on-shell strong-vertex couplings $\CSv,\CDv$, which set the residue of the $d$--$p$ elastic
amplitude at the ${}^{3}\mathrm{He}$ pole through the three-point amplitude $M_{3}=(\CSv\,\mathcal{E}^{(S)}+\CDv\,
\mathcal{E}^{(D)})/m_{d}$ (Eq.~\eqref{eq:strongbasis}). They are fixed by the ANC. The nonrelativistic pole
residue is $(\pi/\mu^{2})\,C_{S}^{2}$. With the relativistic normalization of the
three massive legs restored, the
spin-summed residue reads
\begin{equation}
  |M_{3}|^{2}=\frac{8\pi\,m_{3}m_{d}m_{p}}{\mu^{2}}\,C_{S}^{2}
            \simeq 2.88\times10^{8}\,\mathrm{MeV}^{2} .
  \label{eq:residue}
\end{equation}
We have validated the direct-capture machinery that delivers this
absolute normalization. Evaluated for the deuteron, where the continuum is free of
Coulomb distortion, the same expression reproduces the textbook
Bethe--Peierls~\cite{Bethe:1935diplon} photodisintegration
cross section.

The $d$-wave coupling $C_{D}$ stands on the same footing. Rather than fit it to
the capture data we fix it to the ${}^{3}\mathrm{He}$ asymptotic $D/S$ ratio
$C_{D}/C_{S}=-0.0432$ (Table~\ref{tab:inputs}), computed {\it ab
initio}~\cite{Nollett:2011qf,Kievsky:1997zz} and consistent with the direct
sub-Coulomb transfer measurement of the ${}^{3}\mathrm{He}$
ratio~\cite{Ayer:1995zz} and of the mirror $^3$H ratio~\cite{George:1993zz}. It
enters the $S$-factor only through $|C_{D}|^{2}$, so its sign, which differs by
convention across determinations, never propagates.

\subsubsection{Electric Multipoles: $\eeff$, $q_{\mathrm{eff}}$, $c_{\Eone}$, and $c_{\Etwo}$}

The electric multipoles couple through
recoil effective charges of the relative $d$--$p$ coordinate. The dipole
carries
\begin{equation}
  \eeff = e\,\frac{Z_{d}m_{p}-Z_{p}m_{d}}{m_{d}+m_{p}}
        = e\,\frac{m_{p}-m_{d}}{m_{d}+m_{p}}
        = -0.333\,e,
  \label{eq:eeff}
\end{equation}
nonzero precisely because the deuteron and proton carry different
charge-to-mass ratios.
In the dipole reduced matrix element
$U_{\Eone}=\eeff\,\CSv/m_{d}+(e/\Lambda)\,c_{\Eone}$ [Eq.~\eqref{eq:Udefs}], the point-like piece
$\eeff\,\CSv/m_{d}$ is fixed by the masses and the ANC. The leading structure
contact term $c_{\Eone}$ is then the only free parameter, determined from data in
Sec.~\ref{sec:results}. We quote it through the dimensionless ratio
\begin{equation}
  \tEone \equiv \frac{(e/\Lambda)\,c_{\Eone}}{\eeff\CSv/m_{d}}
        = \frac{e\,m_{d}}{\eeff\,\Lambda}\,\frac{c_{\Eone}}{\CSv}\,,
  \label{eq:tE1def}
\end{equation}
the fractional contact correction to the $\Eone$ amplitude.

The same recoil mechanism fixes the leading electric quadrupole, which
couples through the recoil quadrupole charge of the relative
coordinate,
\begin{equation}
  q_{\mathrm{eff}} = e\,\frac{Z_{d}m_{p}^{2}+Z_{p}m_{d}^{2}}{(m_{d}+m_{p})^{2}}
                   = e\,\frac{m_{p}^{2}+m_{d}^{2}}{(m_{d}+m_{p})^{2}}
                   = +0.555\,e .
  \label{eq:qeff}
\end{equation}
This geometric charge generates a recoil quadrupole $q_{\mathrm{eff}}\langle r^{2}\rangle$.
At the relevant separation $r\sim1/\kappa$ it exceeds the intrinsic deuteron
quadrupole moment $\mathcal{Q}_{d}$ by about an order of magnitude
($\mathcal{Q}_{d}$ enters the amplitude only through the deuteron coupling
$g_{2}^{d}$ of Eq.~\eqref{eq:photond}). The leading $\Etwo$ strength is therefore
set, like the dipole, by the masses and the ${}^{3}\mathrm{He}$ ANC $C_{S}$, with
a few-percent correction from $\mathcal{Q}_{d}$. The analogous $\Etwo$ ratio is
$\tEtwo\equiv e\,m_{d}\,c_{\Etwo}/(q_{\mathrm{eff}}\Lambda\,\CSv)$.
The $d$-wave vertex coupling $\CDv$ rides on the same recoil charge,
$U_{\Etwo}^{(D)}=q_{\mathrm{eff}}\CDv/m_{d}$, but in channel spin $\tfrac32$,
orthogonal to the $S$-state. It therefore adds incoherently rather than shifting
$U_{\Etwo}$. It contributes twice: a constant-in-energy threshold term from the
$^{4}S_{3/2}$ entrance, feeding $a_{0}$ through $\bar A_{\Etwo}^{D}$, and a
$d$-wave admixture from the $^{4}D_{3/2}$ and $^{4}D_{5/2}$ entrances, feeding
$a_{2}$ (Appendix~\ref{app:me}). With the $D/S$ ratio fixed above, the $D$-state effect
on the $\Etwo$ strength is a fixed $(C_{D}/C_{S})^{2}\approx2\times10^{-3}$.

\subsubsection{Magnetic Dipole: $c_{\Mone}$}

The $\Mone$ operator
connects the initial $d+p$ $s$-waves to ${}^{3}\mathrm{He}$ through the total
magnetic moments of the three legs and resolves into two incoherent
channel-spin pieces [Eq.~\eqref{eq:Udefs}]: the $^{2}S_{1/2}$ doublet and a
$^{4}S_{3/2}$ quartet fed only by the proton and deuteron channels. Each carries
its own short-range contact term.

The magnetic dipole is not protected by Siegert's
theorem, and the point-like (leading) piece is subdominant.
Evaluating the cluster ($d\otimes p$) doublet at this point-like
order gives
$\tfrac23\mu_{d}-\tfrac13\mu_{p}=-0.36\,\muN$ for the ${}^{3}\mathrm{He}$ moment, a factor of six below the
measured $\mu_{3}=-2.13\,\muN$. This is an accidental near-cancellation in the one-body moment.
In a potential-model picture the shortfall is supplied by large two-body (meson-exchange) currents, the standard $A=3$ enhancement~\cite{Carlson:1997qn}.
In the EFT language, $c_{\Mone}$ is the coefficient of a dimension-6 contact operator, entering with the explicit factor $e/\Lambda^{3}$ (Table~\ref{tab:couplings})
such that its natural size is $\mathcal{O}(1)$.
Given the cancellation in the leading term, the EFT power counting naturally anticipates
this formally higher-order term will outweigh the leading point-like contribution to the $\Mone$ strength.

The angle-integrated $S$-factor depends on the magnetic dipole through a single
combination---the doublet and quartet strengths added with their channel-spin
weights,
\begin{equation}
  |U_{\Mone}|^{2} \equiv \frac{A_{\Mone}}{\Egam^{2}}
    = \tfrac14\,\big|U_{\Mone}^{(2)}\big|^{2}+\tfrac29\,\big|U_{\Mone}^{(4)}\big|^{2}.
  \label{eq:UM1sq}
\end{equation}
Isolating the doublet and quartet contact terms $c_{\Mone}^{(2)}$ and $c_{\Mone}^{(4)}$
would require an observable sensitive to polarization.
The combination $|U_{\Mone}|^{2}$ is fixed from data in
Sec.~\ref{sec:results}, once the Coulomb-enhanced
electric-dipole tail is removed from the threshold coefficient $a_{0}$. The
anchor is the LUNA measurement~\cite{Casella:2002yej} reaching the
$9\,\mathrm{keV}$ solar Gamow peak. The
resulting short-range strength is corroborated independently in the region
where $c_{\Mone}$ lives. The static ${}^{3}\mathrm{He}$ moment is itself
two-body-enhanced (the standard $A=3$ fingerprint). And the near-threshold
angular distributions of Schmid {\it et al.}~\cite{Schmid:1997zz}
and of LUNA~\cite{Stockel:2024hde} separate the magnetic and electric strengths
directly through their isotropic and $\Mone$--$\Eone$ interference terms.

\subsubsection{Error Budget}
\label{sec:errbudget}

The form of $S(E)$ rests on precisely known inputs, with a single external
quantity dominating the uncertainty. The masses, the $Q$-value, the magnetic
moments, the deuteron quadrupole, and the recoil charges of
Eqs.~\eqref{eq:eeff} and~\eqref{eq:qeff} are all known to far better than the
target precision, so they fix the {\it form} of $S(E)$ essentially exactly.
The leading external uncertainty is the ${}^{3}\mathrm{He}$ asymptotic
normalization $C_{S}$. A rigorous experimental error for the $p+d$ ANC is not
established, so we carry the one-to-two percent model spread discussed above
as the dominant external input. The genuine fit parameters are the
short-range structure contact terms: the leading $\Mone$ and $\Eone$ contact terms
$c_{\Mone},c_{\Eone}$ at dimension $6$, and the subleading
$\Etwo$ contact $c_{\Etwo}$ at dimension $7$ (Appendix~\ref{app:bdy}). With the
explicit factors $e/\Lambda^{n}$ of Eq.~\eqref{eq:Udefs} and
Table~\ref{tab:couplings}, their
naive-dimensional-analysis size is $c_{X}\sim\mathcal{O}(1)$.

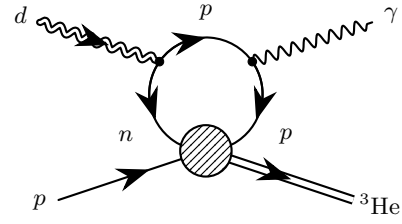
\begin{figure}[t]
\centering
\begin{tikzpicture}[line cap=round]
  \coordinate (V1) at ({0.75*cos(145)},{0.25+0.75*sin(145)});   
  \coordinate (B)  at (0,-0.5);                        
  \coordinate (Pg) at ({0.75*cos(35)},{0.25+0.75*sin(35)});     
  \draw[thick,double,double distance=1.1pt,decorate,
        decoration={snake,amplitude=1.4pt,segment length=6pt}] (-2.2,1.20) -- (V1);
  \draw[draw=none,ar=0.58] (-2.2,1.20) -- (V1);
  \draw[thick] (V1) arc (145:35:0.75);
  \draw[thick,-{Stealth[length=12.5pt,width=9pt]}] (V1) arc (145:90:0.75);
  \draw[thick] (Pg) arc (35:-90:0.75);
  \draw[thick,-{Stealth[length=12.5pt,width=9pt]}] (Pg) arc (35:-27.5:0.75);
  \draw[thick] (V1) arc (145:270:0.75);
  \draw[thick,-{Stealth[length=12.5pt,width=9pt]}] (V1) arc (145:207.5:0.75);
  \draw[thick,ar=0.62] (-1.95,-1.15) -- (B);
  \draw[helion,line cap=butt] (B) -- (1.95,-1.15);
  \draw[thick,decorate,decoration={snake,amplitude=2.2pt,segment length=5pt}]
        (Pg) -- (2.2,1.20);
  \filldraw (V1) circle (1.6pt);
  \filldraw (Pg) circle (1.6pt);
  \fill[white] (B) circle (0.34);
  \fill[pattern=north east lines] (B) circle (0.34);
  \draw[thick] (B) circle (0.34);
  \node at (-2.45,1.30) {$d$};
  \node at (0.00,1.33) {$p$};    
  \node at (-1.06,-0.30) {$n$};  
  \node at (1.06,-0.30) {$p$};   
  \node at (-2.20,-1.18) {$p$};
  \node at (2.45,1.30) {$\gamma$};
  \node at (2.30,-1.20) {${}^{3}\mathrm{He}$};
\end{tikzpicture}
\caption{Representative diagram for the leading EFT contribution from the deuteron breakup to the $\dpHe$ amplitude.}
\label{fig:breakup}
\end{figure}

The theory uncertainty is the EFT truncation, set by the deuteron-breakup cut
(Fig.~\ref{fig:breakup}) that fixes the radius of convergence. Its \emph{strength},
which sets the truncation coefficient, is suppressed near threshold, and naive
dimensional analysis estimates its size. The breakup loop couples through the
deuteron-internal charge dipole $(m_{n}/m_{d})\,e\,\boldsymbol{\rho}$, with
$\boldsymbol{\rho}$ the $n$--$p$ separation. Its coefficient,
$(m_{n}/m_{d})e=0.50\,e$, exceeds the recoil charge $\eeff=-0.33\,e$ that drives
the leading amplitude by the fixed kinematic factor $N\approx1.5$. This is an
NDA-sized two-body current, and it matches the fitted $\tEone$
(Sec.~\ref{sec:results}). Such an NDA
estimate is the expected accuracy for the electroweak contact operators, which in
pionless EFT are not renormalized by the $s$-wave strong
interactions~\cite{Beane:2000fi}. Multiplying $N$ by a single nonrelativistic loop
factor---the two-body unitarity factor $\sim1/\pi$ times an $\mathcal{O}(1)$
threshold phase space---gives the coefficient $R\sim N/8\approx0.2$ in the
truncation term $R\,(\prel/\pstar)^{2}$. This is several times below the
$\mathcal{O}(1)$ a generic contact term would carry, and the slowly opening phase
space and the $pp$ Coulomb repulsion push it lower still. The
truncation is thus nominally $\sim\!1\%$ at the Gamow peak, rising to $\sim\!5\%$ at
the top of the window, and is bounded directly by the data (Sec.~\ref{sec:results}).

\subsection{Comparison to Existing Calculations and Data}
\label{sec:comparison}

The structure of Eq.~\eqref{eq:SE} matches the established picture of the
$\dpHe$ $S$-factor, and the on-shell decomposition maps cleanly onto the {\it ab
initio} one. The benchmark is set by two hyperspherical-harmonics calculations of
Marcucci {\it et al.}, both AV18/UIX. The 2005 analysis~\cite{Marcucci:2005zc}
supplies the energy-dependent multipole currents and the one-body/two-body
decomposition we map onto. The 2016 calculation~\cite{Marcucci:2015yla} provides
the BBN-window $S$-factor, the curve adopted by Solar Fusion~III~\cite{Acharya:2024lke}. The {\it ab initio} curve reproduces
the \emph{shape} of the data well but sits high in absolute normalization: the precise
LUNA points lie up to $\sim\!10\%$ below it across the BBN window. The Solar
Fusion~III Bayesian analysis quantifies the offset: fitting a rescaled version
of the 2016 curve to the data returns a scale factor $a=0.921(19)$, an
overestimate of $7.9\%$~\cite{Acharya:2024lke}. The curve's quoted percent-level accuracy thus reflects
internal convergence, not agreement with the post-LUNA data. Earlier nucleon-level
EFT~\cite{Sadeghi:2008zz} and potential cluster-model calculations reproduce the
threshold behavior, and the network analyses of Refs.~\cite{Yeh:2020mgl,Moscoso:2021xog}
provide the Bayesian rate evaluations the BBN codes adopt.

EFT and {\it ab initio} can be compared only through the same per-multipole
totals. What the angle-integrated $S$-factor cannot do is resolve each total
into its strong and electromagnetic (one- versus two-body) parts. The {\it ab initio}
calculation quotes that resolution, but supplies it from its interaction and current
operators rather than from the data: the split is defined relative to that scheme, and a
different Hamiltonian apportions the same total differently. The EFT makes the shared
constraint explicit.

The experimental landscape is well suited to fixing the few free parameters. The fits
below combine the LUNA~\cite{Casella:2002yej,Mossa:2020gjc} and
T\"{u}rkat~\cite{Turkat:2021qmq} measurements, which together span the solar Gamow peak
through the BBN window (Appendix~\ref{app:efffit}). The Ti\v{s}ma~\cite{Tisma:2019acf}
and older Schmid~\cite{Schmid:1997zz} measurements are consistent with this picture but,
with larger or partly correlated uncertainties, are not included in the quantitative fit.
Because the threshold and the window constrain different combinations of the magnetic and
electric strengths (Sec.~\ref{sec:sfactor}), a global analysis must combine both energy
ranges to separate them.

\section{Results}
\label{sec:results}

We determine the effective-field-theory parameters by a single global fit to
the capture data and the theory inputs. The $S$-factor of Eq.~\eqref{eq:SE}
is evaluated directly from the fit parameters: the ${}^{3}\mathrm{He}$ ANC
$C_{S}$, the magnetic strength $|U_{\Mone}|^{2}$, the electromagnetic contact
terms $c_{\Eone}$, $c_{\Etwo}$, and $c_{M2}$, and one overall normalization
$\lambda_{e}$ per experiment. The posterior adds two theory priors to the data
$\chi^{2}$: the ${}^{3}\mathrm{He}$ asymptotic normalization as a Gaussian
prior on $C_{S}$, and naive-dimensional-analysis priors on the short-range
contact terms. Each experiment's quoted scale uncertainty enters through
$\lambda_{e}$ (Casella $5\%$, Mossa $2.5\%$, T\"{u}rkat $12\%$). The EFT
truncation enters as a per-point theory error $R\,(\prel/\pstar)^{2}$, with
$\pstar$ the breakup scale that sets the radius of convergence of the energy
expansion (Sec.~\ref{sec:eft}) and $R\sim N/8$ the NDA shape coefficient of
Sec.~\ref{sec:errbudget}. Up to a parameter-independent constant, the negative
log-posterior is
\begin{equation}
\begin{aligned}
  \chi^{2}\!\left(\theta,\{\lambda_{e}\}\right) =\;
  & \sum_{e}\sum_{i\in e}
    \frac{\big[\,S_{i}^{(e)}-\lambda_{e}\,S(E_{i};\theta)\,\big]^{2}}{\big(\sigma_{i}^{(e)}\big)^{2}} \\
  & +\left(\frac{C_{S}-\bar C_{S}}{\sigma_{C_{S}}}\right)^{2}
    +\sum_{X}\left(\frac{c_{X}}{\sigma_{X}}\right)^{2} \\
  & +\sum_{e}\left(\frac{\lambda_{e}-1}{\delta_{e}}\right)^{2},
\end{aligned}
  \label{eq:chi2}
\end{equation}
where the per-point variance
$\big(\sigma_{i}^{(e)}\big)^{2}=\big(s_{i}^{(e)}\big)^{2}+\big[\,R\,S_{i}^{(e)}(\prel^{(i)}/\pstar)^{2}\,\big]^{2}$
combines the per-point (uncorrelated) error $s_{i}^{(e)}$ and the EFT truncation. Here
$\theta=\{|U_{\Mone}|^{2},C_{S},c_{\Eone},c_{\Etwo},c_{M2}\}$ are the fit
parameters (the electric strengths built from $\CSv(C_{S})$ via the residue relation of
Sec.~\ref{sec:matching}), $\bar C_{S}=2.144\,\mathrm{fm}^{-1/2}$ and
$\sigma_{C_{S}}=0.02\,\bar C_{S}$ the ANC prior, $\sigma_{X}$ the
naive-dimensional-analysis widths on the coefficients $X\in\{\Eone,\Etwo,M2\}$, and
$\delta_{e}$ the per-experiment scale uncertainties above. The width on
$c_{\Eone}$ is left wide, so $c_{\Eone}$ follows from the data and the ANC
prior; the magnetic strength carries no prior. The posterior
$\propto e^{-\chi^{2}/2}$ is explored by Markov-chain Monte Carlo.

\begin{figure}[t]
  \centering
  \includegraphics[width=\columnwidth]{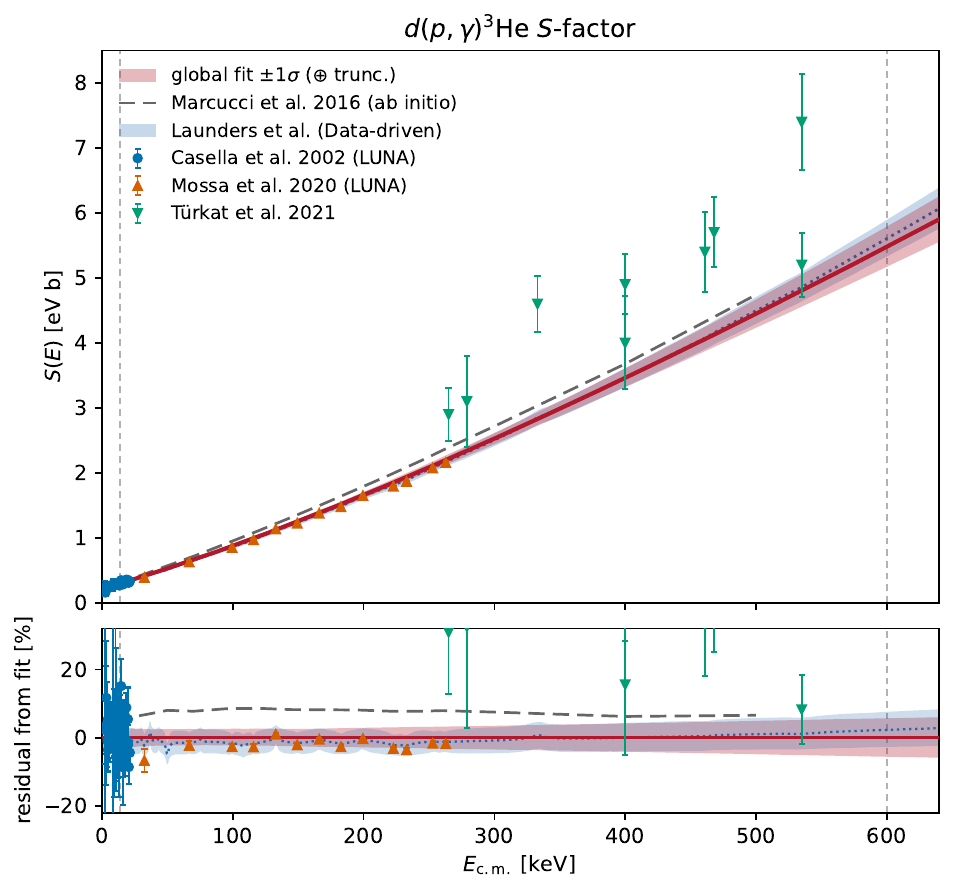}
  \caption{The $\dpHe$ $S$-factor from the global Bayesian fit within the BBN window: posterior band ($\pm1\sigma$ on the fit parameters
  $\oplus$ the $R(\prel/\pstar)^{2}$ truncation) through the LUNA
  data~\cite{Casella:2002yej,Mossa:2020gjc}, with the {\it ab initio} Marcucci
  curve~\cite{Marcucci:2015yla} (dashed) and the Gaussian-process band of Launders
  {\it et al.}~\cite{Launders:2026ciu} (blue, agreeing with the fit across the window)
  overlaid; the T\"{u}rkat data~\cite{Turkat:2021qmq} sit high and are
  included with a floated normalization ($\lambda=1.28$), so they do not set
  the absolute scale.
  Lower panel: per-point residuals.}
  \label{fig:global}
\end{figure}

The result is shown in Fig.~\ref{fig:global} and collected in Table~\ref{tab:lecs}. The
global fit gives $\chi^{2}/\mathrm{dof}=0.83$, just below unity and consistent
with the truncation estimate of Sec.~\ref{sec:errbudget}. The band passes
through the LUNA data, with an extrapolated threshold value
$S(0)=0.209\pm0.008\,\mathrm{eV\,b}$, between the Solar
Fusion~III recommendation ($0.203\pm0.005\,\mathrm{eV\,b}$~\cite{Acharya:2024lke}) and the Marcucci
benchmark ($0.216\,\mathrm{eV\,b}$~\cite{Marcucci:2015yla}), and consistent with both.
The higher-energy T\"{u}rkat data sit systematically above the LUNA-anchored
curve: the fitted normalization $\lambda_{\text{T\"urkat}}=1.28$ lies
$\sim\!2.3\sigma$ beyond the quoted $12\%$ scale.  Similar T\"{u}rkat--LUNA tension is also reported by
the data-driven Gaussian-process analysis of Ref.~\cite{Launders:2026ciu}.

For analytic applications, the $\pm1\sigma$ band about the central $S(E)$ of
Sec.~\ref{sec:dpHe} is well described across the BBN window by the fractional
half-width
\begin{equation}
  \frac{\Delta S(E)}{S(E)} \simeq \sqrt{\sigma_{0}^{2}+\big(R\,E/B_{d}\big)^{2}},
  \qquad \sigma_{0}\simeq0.024,
  \label{eq:band}
\end{equation}
where the data-and-ANC floor $\sigma_{0}$ is added in quadrature with the EFT truncation
($R\simeq0.19$, $B_{d}=2.2246\,\mathrm{MeV}$).  The band edges are at $S(E)\,[1\pm\Delta S/S]$.
Equation~\eqref{eq:band} reproduces the band of Fig.~\ref{fig:global} to better
than $0.3$ percentage points from the Gamow peak to the top of the window.
Below the Gamow peak it underestimates the width: as the magnetic dipole comes
to dominate, the exact band widens to $\Delta S(0)/S(0)\simeq0.04$ at
threshold. The exact band accompanies the paper as ancillary data.

\begin{table}[t]
\caption{EFT parameters from the global fit: posterior median
and $68\%$ interval.  The electric dipole
and the magnetic strength are data-determined, whereas the quadrupole contact terms are
prior-limited.}
\label{tab:lecs}
\begin{ruledtabular}
\begin{tabular}{l l}
Quantity & ~~~~~~~Value \\
\hline
$C_{S}$ (ANC)                              & $2.14\pm0.04\,\mathrm{fm}^{-1/2}$ \\
$\tEone$                                   & $-0.146\pm0.021$                  \\
$|1+\tEone|$                               & $0.85\pm0.02$                     \\
$C_{S}^{\rm eff}=C_{S}(1{+}\tEone)$        & $1.83\pm0.03\,\mathrm{fm}^{-1/2}$  \\
$S_{\Mone}(0)$                             & $0.107\pm0.008\,\mathrm{eV\,b}$   \\
$\tEtwo$                                   & $0.0\pm0.5$                       \\
$c_{M2}$                                   & consistent with $0$              \\
$\lambda$ (Casella, Mossa, T\"{u}rkat)         & $1.01(3),\ 0.98(2),\ 1.28(6)$     \\
\hline
$S(0)$                                     & $0.209\pm0.008\,\mathrm{eV\,b}$   \\
$\chi^{2}/\mathrm{dof}$                    & $0.83$ \ ($N=73$)                 \\
\end{tabular}
\end{ruledtabular}
\end{table}

The leading contribution to the electric-dipole sector is the point-like piece
$\eeff\CSv/m_{d}$, with the vertex coupling fixed by the ${}^{3}\mathrm{He}$ ANC through the
residue relation of Sec.~\ref{sec:matching}.  It is corrected by the next-to-leading contact term $c_{\Eone}$,
with the fit returning,
\begin{equation}
  \tEone = -0.146\pm0.021,\qquad
  |1+\tEone| = 0.85,
  \label{eq:cE1fit}
\end{equation}
a $\sim\!15\%$ deficit in the $\Eone$ amplitude ($\sim\!30\%$ in the rate).
The {\it ab initio} Marcucci curve, carrying its own two-body
currents, falls between the point-like ($c_{\Eone}=0$) prediction and the data (Sec.~\ref{sec:comparison}).
The fit to the capture data constrains the electric-dipole amplitude, fixing the product
\begin{equation}
  C_{S}^{\rm eff}\equiv C_{S}\big(1+\tEone\big)=1.83\,\mathrm{fm}^{-1/2} .
  \label{eq:CSeff}
\end{equation}
The ${}^{3}\mathrm{He}$ ANC theory prior disentangles $C_S$ from $c_{\Eone}$.
The resulting uncertainty of $\pm0.021$ on $\tEone$ is set mainly
by the $\sim\!2\%$ ANC prior, with the remainder from the truncation-inflated
data errors.

The fitted $\tEone$ is of natural size, a non-trivial outcome given that its prior was left wide.
The naive-dimensional-analysis scale of Sec.~\ref{sec:eft}
($\Lambda\sim\sqrt{4\pi}\,\pstar\approx187\,\mathrm{MeV}$) places a natural
$\tEone$ at $\mathcal{O}(10\text{--}40\%)$, the upper end set by
$\kappa/\Lambda\approx0.4$. The fitted $-0.15$ sits comfortably inside, and is
the same size as the $\sim\!10\%$ two-body-current shift the {\it ab initio}
calculation carries in this channel (Sec.~\ref{sec:comparison}).
The contrast with the magnetic dipole is instructive. $c_{\Mone}$ and
$c_{\Eone}$ are operators of the same order---the leading dimension-6 contact
terms---but the point-like magnetic moment nearly cancels
(Sec.~\ref{sec:matching}), so $c_{\Mone}$ dominates the magnetic strength and
carries the bulk of the magnetic share of $S(0)$. The unsuppressed Siegert
convection current instead leaves $c_{\Eone}$ a small correction.

The fact that the offset between {\it ab initio} theory and data lives in the electric-dipole channel, not in a global scale error or a
higher-order truncation, is established by two observations. It has the wrong shape to be
truncation: the omitted orders grow as $(\prel/\pstar)^{2}$ (Sec.~\ref{sec:eft}), whereas
the offset is already $\sim\!10\%$ at low energy, and a free cubic term in the effective
fit (Appendix~\ref{app:efffit}) comes back consistent with zero. And it is
$\Eone$-localized. Any $\Eone$ effect must track the Coulomb-aware
electric-dipole fraction $f_{\Eone}(E)=S_{\Eone}/(S_{\Mone}+S_{\Eone})$, which
runs from $\approx0.5$ at threshold to $\approx0.97$ at the top of the window.
A global scale error is instead flat in $f_{\Eone}$. Fitting the deficit
against $f_{\Eone}$ favors the localized form over a constant by
$\Delta\chi^{2}\approx9$. What this does \emph{not} decide is the offset's
nature within the channel: a short-range two-body current $c_{\Eone}$ and a
correspondingly smaller effective ANC [Eq.~\eqref{eq:CSeff}] are
indistinguishable to the angle-integrated $S$-factor.

The quadrupole contact terms are prior-limited. The angle-integrated $S$-factor
constrains only three combinations (the $a_{0},a_{1},a_{2}$ of the effective
fit), and the genuine quadrupole content of $a_{2}$ sits at the percent level
(Sec.~\ref{sec:sfactor}). Once the dipoles are fixed, $c_{\Etwo}$ and the $M2$
strength (floated as a single effective contact) are therefore in practice
unconstrained: their posteriors return the naive-dimensional-analysis priors
(Table~\ref{tab:lecs}). Sitting one order in the expansion below the present
sensitivity, they leave the electric-dipole determination and $S(0)$ unchanged
whether floated or held fixed.

\section{Discussion and Outlook}
\label{sec:outlook}

We have constructed proton--deuteron radiative capture as a nuclear-state
effective field theory, with the deuteron, proton, and ${}^{3}\mathrm{He}$ treated as
point-like and the amplitude assembled using on-shell methods. Every tree-level
structure consistent with Lorentz invariance, little-group covariance, and the
electromagnetic Ward identity is enumerated, including the boundary
(two-body-current) terms. The gauge-invariant $S$-factor then reduces to an
incoherent multipole tower whose strengths are explicit functions of measured
electromagnetic moments, the ${}^{3}\mathrm{He}$ asymptotic normalization, and
a handful of short-range contact terms. The parameter-to-observable map is transparent: the threshold $S(0)$ is
shared between the magnetic dipole and the Coulomb-enhanced electric-dipole tail, the
window-wide rise is the electric dipole, and the curvature dominated by the photon phase space.

The electric dipole is the cleanest result. Its \emph{leading} strength is fixed
by the exact recoil charge and the ${}^{3}\mathrm{He}$ ANC. This leading EFT
prediction lies above the precise LUNA data by an amount whose size and energy
dependence are just what the power counting assigns to the next-to-leading
$\Eone$ correction. A single natural contact term,
$\tEone\simeq-0.15$, absorbs the offset. The magnetic dipole is
the mirror case: its point-like (impulse) moment has a strong accidental
cancellation, and the same-order contact term dominates
(Sec.~\ref{sec:matching}). Both the large threshold $\Mone$ and the small
$\Eone$ correction thus trace to a single feature---whether the leading term is
suppressed---and the short-range sector is free of fine-tuning.

Two limitations bound the present analysis, and each points to its own next step.
The capture data fix only the total $\Eone$ amplitude
$U_{\Eone}=\eeff\CSv/m_{d}+(e/\Lambda)c_{\Eone}$; separating the
contact term from a correspondingly smaller effective ANC requires a second observable.
The most direct is a precise low-energy $p$--$d$ doublet (${}^{2}S_{1/2}$)
elastic-scattering analysis: the Coulomb-modified effective-range function
carries the ANC as the residue at the ${}^{3}\mathrm{He}$ pole. This is the
same framework that supplies the {\it ab initio} value~\cite{Kievsky:1997zz}
and that pionless-EFT analyses of $pd$ scattering and ${}^{3}\mathrm{He}$ also
provide~\cite{Vanasse:2014kxa,Konig:2015aka}, though the near-threshold doublet
pole makes the extraction delicate. A modern sub-Coulomb transfer measurement
is the alternative. The rigorous resolution is a joint fit of capture and
elastic data with the ANC shared. Even then $c_{\Eone}$ remains one constant:
its decomposition into continuum distortion and exchange current is defined by
a scheme rather than by an observable. First-principles
input to the analogous short-range two-body currents is beginning to emerge from lattice
QCD~\cite{Savage:2016kon,Davoudi:2020ngi}, offering a future route to fixing such constants
independently of the capture data.
Separately, the angle-integrated $S$-factor constrains the magnetic dipole only
through the summed strength of Eq.~\eqref{eq:UM1sq}, and the quadrupoles only
through a single combination. The individual doublet and quartet $c_{\Mone}$
and the quadrupole contact terms are therefore degenerate here. They would be
split only by angular or polarization observables, as in the chiral-interaction
treatment of nucleon--deuteron capture analyzing
powers~\cite{Skibinski:2006gy}. The photon angular distributions now measured
by LUNA across the BBN window~\cite{Stockel:2024hde} are the nearest such
handle, and folding them into the fit is a natural extension.

For Big Bang nucleosynthesis the relevant comparison is to the $S$-factors the
network codes adopt~\cite{Gariazzo:2021iiu,Pitrou:2018cgg,Moscoso:2021xog,Burns:2023sgx,Giovanetti:2024zce}. At the BBN Gamow peak the
data-driven fit of Appendix~\ref{app:efffit} agrees with the polynomial fit used
by PArthENoPE~\cite{Gariazzo:2021iiu} and with the Gaussian-process posterior of
Ref.~\cite{Launders:2026ciu} to better than one percent, while it sits
$\sim\!7\%$ below the scaled {\it ab initio} curve adopted in
PRIMAT~\cite{Pitrou:2018cgg,Moscoso:2021xog}. Across the window the PArthENoPE
agreement stays at the percent level, the fit remains inside the
Gaussian-process band, and the PRIMAT offset ranges over $5$--$8\%$.

This is the central phenomenological result of this work: the
PRIMAT--PArthENoPE difference is recast from a choice between curves into a
measurable question. The same theory-versus-data offset carried by the global
fit of Fig.~\ref{fig:global} is traced to a single low-energy constant rather
than an unresolved normalization---a correction the power counting anticipates
at next-to-leading order (Sec.~\ref{sec:results}). The elastic $d$--$p$
observable above can determine that constant independently of the capture
data. The deuterium abundance the network infers inherits this difference, so
its origin can be settled experimentally rather than adjudicated between
curves. This is of direct relevance to the
$\mathrm{D}/\mathrm{H}$ tension~\cite{Launders:2026ciu}.

The obvious continuation is to propagate $S(E)$ and its covariance through the BBN
network, and to apply the same on-shell construction to the transfer reactions
$d(d,n){}^{3}\mathrm{He}$ and $d(d,p){}^{3}\mathrm{H}$. There the amplitude is
genuinely richer: a tower of ${}^{4}\mathrm{He}$ compound states saturates the
$s$-channel in place of a single bound-state pole, and the identical-deuteron
entrance enforces Bose symmetry. The mirror exit channels couple through
isospin, with Coulomb distortion in the $d(d,p){}^{3}\mathrm{H}$ exit. And
with no photon, no coupling is Ward-locked, so the contact layer carries more
freedom. But every ingredient is the same kind of object as
here: on-shell vertices glued on physical poles, completed by an enumerated
contact tower and matched to data. These reactions carry the dominant nuclear
uncertainty in the post-LUNA $\mathrm{D}/\mathrm{H}$
budget~\cite{Pitrou:2020etk,Pisanti:2020efz}. Unlike $\dpHe$, which is well
measured and well calculated, they lack both a percent-level {\it ab initio}
benchmark and a complete data-driven description, so a systematic EFT treatment
with quantified truncation errors is most needed there. We defer their
implications for the predicted $\mathrm{D}/\mathrm{H}$ to future work. On the
experimental side, the upper half of the BBN window
($300$--$600\,\mathrm{keV}$), where the T\"urkat normalization sits
$2.3\sigma$ high (Sec.~\ref{sec:results}), has no deep-underground
measurement. A LUNA-MV-class extension of the underground program to those
energies would settle the normalization directly. More broadly, the present
calculation opens the way to a broader exploration of amplitude-derived
effective field theories for nuclear reactions.

\section*{Acknowledgements}

The author is grateful for illuminating conversations with Vincenzo Cirigliano, Susan Gardner, Shirley Li, Toni M\"{a}kel\"{a}, George Parker, 
Martin Savage, Adrian Thompson, Mauro Valli, and especially Ryan Plestid and Cara Giovanetti.
The research of TMPT is supported in part by the National Science Foundation
through Grant PHY-2514888.  Feynman diagrams were produced with the Feynmangraphz package by Michael Ratz.\\

\noindent
{\bf AI Usage:}~All analytic calculations and numerical results were primarily derived using Claude Opus 4.8 by Anthropic, guided and cross-checked/validated by
the human author, who takes full responsibility for all of the results presented here.

\appendix

\section{On-Shell Construction of the Capture Amplitude}
\label{app:amp}

This appendix records the analytic construction of the $\dpHe$ amplitude.
We use the massive
spinor-helicity conventions of Ref.~\cite{Arkani-Hamed:2017jhn}: bold
spinors $|\bm{i}^{I}\rangle,|\bm{i}^{I}]$ carry the $SU(2)$ little-group
index $I$ of a massive leg $i$. Amplitudes are polynomials in the
angle and square brackets $\langle\,\rangle,[\,]$. All legs are taken
incoming, so the physical capture has the deuteron and proton as particles
and ${}^{3}\mathrm{He}$ entering as its antiparticle $\Hb$. The four
contributions to the amplitude, collected in Fig.~\ref{fig:diagrams}, are constructed below.

\subsection{Three-Point Vertices}
\label{app:vertices}

The Wilson coefficients are collected in
Table~\ref{tab:couplings}, including the three-point vertex couplings which assemble into four-point amplitudes via factorization, and
the four-point boundary terms. Each entry multiplies the listed bracket
structure, and the equations below introduce them in that same notation.

\begin{table*}[t]
\caption{Wilson coefficients entering the on-shell $\dpHe$ amplitude. \emph{Above the
rule:} three-point vertex couplings; the photon-vertex structures are written with
generic massive lines $\bm{a},\bm{b}$ on the radiating leg and the photon $\gamma$.
\emph{Below the rule:} four-point boundary currents with legs
$(1,2,3,4)=(\Hb,d,p,\gamma)$. Bold spinors carry the symmetrized little-group
indices of a massive leg.
Each coupling entry is the full coefficient of the listed bracket structure:
a dimensionless coupling times its explicit normalization (charges $e$,
moments $e/2m_{X}$, spin-$1$ polarization $1/m_{d}$, contact terms $e/\Lambda^{3}$
with $\Lambda=\sqrt{4\pi}\,\pstar$). The last column is the operator dimension in the standard relativistic
counting (fermion $\tfrac32$, boson and derivative $1$, photon field strength $2$), in
which the leading two-body currents enter at dimension $6$ (Sec.~\ref{sec:matching}).
The operator dimension is not fixed by the bracket normalization: on the
three-point locus derivatives evaluate to masses, so, e.g., the dimension-$4$
and dimension-$5$ strong operators share one on-shell coefficient form, the
derivative content of $\Ctwo$ surfacing instead in the Ward term of
Eq.~\eqref{eq:BWbracket}.}
\label{tab:couplings}
\begin{ruledtabular}
\begin{tabular}{l l l c}
Coupling & Amplitude Structure & Physical Interpretation & Dim \\
\hline
$\Cone/m_{d}$ & $\ad{\bm{d}}{\bm{p}}\ad{\bm{d}}{\bm{h}}+\sq{\bm{d}}{\bm{p}}\sq{\bm{d}}{\bm{h}}$ & strong, vector & $4$ \\
$\Ctwo/m_{d}$ & $\ad{\bm{d}}{\bm{p}}\sq{\bm{d}}{\bm{h}}+\sq{\bm{d}}{\bm{p}}\ad{\bm{d}}{\bm{h}}$ & strong, tensor & $5$ \\
$Z_{p}e$      & $x\,\ad{\bm{a}}{\bm{b}}$                                     & proton charge            & $4$ \\
$e\,\kappa_{p}/2m_{p}$   & $\sq{\bm{a}}{\gamma}\sq{\bm{b}}{\gamma}$                     & proton moment & $5$ \\
$Z_{3}e$    & $x\,\ad{\bm{a}}{\bm{b}}$                                     & ${}^{3}$He charge        & $4$ \\
$e\,\kappa_{3}/2m_{3}$ & $\sq{\bm{a}}{\gamma}\sq{\bm{b}}{\gamma}$                     & ${}^{3}$He moment        & $5$ \\
$Z_{d}e/m_{d}$      & $x\,\ad{\bm{a}}{\bm{b}}^{2}$                                 & deuteron charge          & $4$ \\
$e\,g_{1}^{d}/m_{d}^{2}$   & $x^{2}\ad{\gamma}{\bm{a}}\ad{\gamma}{\bm{b}}\ad{\bm{a}}{\bm{b}}$ & deuteron $\mu_{d}$    & $4$ \\
$e\,g_{2}^{d}/m_{d}^{3}$   & $x^{3}\ad{\gamma}{\bm{a}}^{2}\ad{\gamma}{\bm{b}}^{2}$        & deuteron quadrupole      & $5$ \\
\hline
$\mathcal{B}_{W}$ & Eq.~\eqref{eq:BWbracket} & Ward (gauge) & $5$ \\
$(e/\Lambda^{3})\,c_{1}$ & $\ad{\bm{1}}{\bm{2}}\ad{\bm{2}}{4}\ad{\bm{3}}{4}$ & 2-body current      & $6$ \\
$(e/\Lambda^{3})\,c_{2}$ & $\ad{\bm{1}}{4}\ad{\bm{2}}{\bm{3}}\ad{\bm{2}}{4}$ & 2-body current      & $6$ \\
$(e/\Lambda^{3})\,c_{3}$ & $\ad{\bm{1}}{4}\ad{\bm{2}}{4}\sq{\bm{2}}{\bm{3}}$ & 2-body current      & $6$ \\
$(e/\Lambda^{3})\,c_{4}$ & $\ad{\bm{2}}{4}\ad{\bm{3}}{4}\sq{\bm{1}}{\bm{2}}$ & 2-body current      & $6$ \\
$(e/\Lambda^{3})\,c_{5}$ & $\ad{\bm{2}}{4}^{2}\sq{\bm{1}}{\bm{3}}$           & 2-body current ($M1$) & $6$ \\
\end{tabular}
\end{ruledtabular}
\end{table*}

\subsubsection{Strong Vertex}

The $d$--$p$--${}^{3}\mathrm{He}$ coupling is a three-massive-leg
vertex. The naive counting gives
$(2S_{d}+1)(2S_{p}+1)(2S_{3}+1)=12$ structures, but on the three-particle
locus momentum conservation and the mass-shell relations collapse these to
four independent bracket monomials. Equivalently, the Jacob--Wick helicity
count for $\tfrac12\to1\otimes\tfrac12$, with $|\lambda_{d}-\lambda_{p}|\le\tfrac12$,
allows exactly four amplitudes. Imposing parity conservation leaves two independent
parity-even couplings. Suppressing and symmetrizing over the deuteron
little-group indices, these are the chirality-diagonal and chirality-mixed
structures
\begin{equation}
\begin{split}
  \mathcal{E}^{(1)} &= \ad{\bm{d}}{\bm{p}}\ad{\bm{d}}{\bm{h}} + \sq{\bm{d}}{\bm{p}}\sq{\bm{d}}{\bm{h}},\\
  \mathcal{E}^{(2)} &= \ad{\bm{d}}{\bm{p}}\sq{\bm{d}}{\bm{h}} + \sq{\bm{d}}{\bm{p}}\ad{\bm{d}}{\bm{h}},
\end{split}
  \label{eq:strongbasis}
\end{equation}
so that the strong vertex is
$M_{3} = \big(\Cone\mathcal{E}^{(1)}+\Ctwo\mathcal{E}^{(2)}\big)/m_{d}$, with
dimensionless couplings and the spin-$1$ polarization factor $1/m_{d}$
explicit.
Diagonalizing the $d$--$p$ channel spin, the chirality combinations:
\begin{equation}
\begin{split}
  \mathcal{E}^{(S)} &= \mathcal{E}^{(1)}+\mathcal{E}^{(2)}
     = \big(\ad{\bm{d}}{\bm{p}}+\sq{\bm{d}}{\bm{p}}\big)
       \big(\ad{\bm{d}}{\bm{h}}+\sq{\bm{d}}{\bm{h}}\big),\\
  \mathcal{E}^{(D)} &= \mathcal{E}^{(1)}-\mathcal{E}^{(2)}
     = \big(\ad{\bm{d}}{\bm{p}}-\sq{\bm{d}}{\bm{p}}\big)
       \big(\ad{\bm{d}}{\bm{h}}-\sq{\bm{d}}{\bm{h}}\big)
\end{split}
  \label{eq:strongdiag}
\end{equation}
separate the channel-spin content. $\mathcal{E}^{(S)}$ is purely the
$(L,s)=(0,\tfrac12)$ $^{2}S_{1/2}$ doublet. $\mathcal{E}^{(D)}$ is
dominantly the $(L,s)=(2,\tfrac32)$ channel-spin-$\tfrac32$ quartet, the
${}^{4}D_{1/2}$ $D$-state of ${}^{3}\mathrm{He}$ (weight $\tfrac89$; the
$\tfrac19$ doublet remainder enters observables only at higher order in
$\prel$). These are the two parity-even channels reaching
$J^{\pi}=\tfrac12^{+}$. In this basis
\begin{equation}
\begin{gathered}
  M_{3} = \frac{1}{m_{d}}\Big(\CSv\,\mathcal{E}^{(S)} + \CDv\,\mathcal{E}^{(D)}\Big),\\
  \CSv=\tfrac12(\Cone+\Ctwo),\quad \CDv=\tfrac12(\Cone-\Ctwo),
\end{gathered}
  \label{eq:CSCD}
\end{equation}
with $\CSv$ the $s$-wave and $\CDv$ the $d$-wave vertex coupling;
we use the diagonal basis $\{\CSv,\CDv\}$ in the reduced matrix elements below.

\subsubsection{Photon Interactions}

The photon couples to each charged leg through a vertex with two equal-mass lines
$\bm{a},\bm{b}$ and the photon $\gamma$. The proton and ${}^{3}\mathrm{He}$
vertices are morphologically identical but with distinct couplings:
\begin{eqnarray}
  M^{+1}_{p,\,IJ}   &=& Z_{p}e\,x\,\ad{\bm{a}^{I}}{\bm{b}^{J}}
                      + \frac{e\,\kappa_{p}}{2m_{p}}\,\sq{\bm{a}^{I}}{\gamma}\sq{\bm{b}^{J}}{\gamma},
                      \label{eq:photonp}\\
  M^{+1}_{3,\,IJ} &=& Z_{3}e\,x\,\ad{\bm{a}^{I}}{\bm{b}^{J}}
                      + \frac{e\,\kappa_{3}}{2\mh}\,\sq{\bm{a}^{I}}{\gamma}\sq{\bm{b}^{J}}{\gamma},
                      \label{eq:photonHe}
\end{eqnarray}
with the negative-helicity image obtained by $\langle\,\rangle\!\leftrightarrow\![\,]$
and $x\to1/x$, where $x$ is the standard massless-leg factor. The charges
$Z_{p}=+1$ and $Z_{3}=+2$ are fixed by the Ward identity. The dimensionless
anomalous moments $\kappa_{X}$ are free LECs and drive the magnetic-dipole transition. The physical
quantity entering that transition is the total magnetic moment: for the
spin-$\tfrac12$ legs the Gordon decomposition splits it into the Ward-fixed
Dirac piece plus the anomalous LEC,
\begin{equation}
  \mu_{X} = \big(Z_{X}+\kappa_{X}\big)\frac{m_{p}}{m_{X}}\quad[\mu_{N}],
  \qquad \mu_{N}=\frac{e}{2m_{p}},
  \label{eq:totalmoment}
\end{equation}
e.g.\ $\mu_{p}=Z_{p}+\kappa_{p}=2.793$. For the spin-$1$ deuteron, whose
minimal coupling carries $g=2$, Eq.~\eqref{eq:totalmoment} instead simply
defines $\kappa_{d}$, and the dimensionless $g_{1}^{d}$ is matched to the total $\mu_{d}$. It is
these $\mu_{X}$, tabulated in
Table~\ref{tab:inputs}, that appear in the magnetic-dipole reduced matrix element. The deuteron, a
charged spin-$1$ field, carries three form factors (little-group indices
suppressed),
\begin{eqnarray}
  M^{+1}_{d} &=& \frac{Z_{d}e}{m_{d}}\,x\,\ad{\bm{a}}{\bm{b}}^{2}
         + \frac{e\,g_{1}^{d}}{m_{d}^{2}}\,x^{2}\,\ad{\gamma}{\bm{a}}\ad{\gamma}{\bm{b}}\ad{\bm{a}}{\bm{b}}
         \nonumber\\
         && +\, \frac{e\,g_{2}^{d}}{m_{d}^{3}}\,x^{3}\,\ad{\gamma}{\bm{a}}^{2}\ad{\gamma}{\bm{b}}^{2},
  \label{eq:photond}
\end{eqnarray}
with charge $Z_{d}=+1$, the dimensionless magnetic dipole $g_{1}^{d}$
fixed by the deuteron moment $\mu_{d}$, and the dimensionless electric quadrupole $g_{2}^{d}$ fixed
by the deuteron quadrupole \emph{moment} $\mathcal{Q}_{d}=+0.286\,e\,\mathrm{fm}^{2}$
(Table~\ref{tab:inputs}).

\subsection{Factorized Four-Point Amplitude and the Ward Identity}
\label{app:fact}

On-shell recursion reconstructs the factorizable part of the capture
amplitude from its residues on the three physical poles, one per charged leg
radiating the photon [Fig.~\ref{fig:diagrams}(a)--(c)]:
internal ${}^{3}\mathrm{He}$ ($s_{dp}=\mh^{2}$, final-state radiation),
internal proton ($s_{p\gamma}=m_{p}^{2}$), and internal deuteron
($s_{d\gamma}=m_{d}^{2}$). Each channel glues the strong vertex of
Eq.~\eqref{eq:strongbasis} to a photon vertex across the internal line, with
massless-leg factors
\begin{equation}
  x_{3}=-\frac{\sq{h}{\gamma}}{\ad{h}{\gamma}},\quad
  x_{p}=+\frac{\sq{p}{\gamma}}{\ad{p}{\gamma}},\quad
  x_{d}=+\frac{\sq{d}{\gamma}}{\ad{d}{\gamma}}.
\end{equation}
The relative sign distinguishes final- from initial-state radiation.

The factorized piece is not gauge invariant on its own. Replacing the photon
polarization by its momentum, $\varepsilon_{\mu}(\gamma)\to k_{\mu}$, the
convection ($g_{0}$) part of each channel collapses against its propagator,
and the charges sum to
\begin{eqnarray}
  k_{\mu}M^{\mu}_{\mathrm{fact}}\big|_{\mathrm{charge}}
  &=& e\,(Z_{d}+Z_{p}+Z_{\bar 3})\,V_{\mathrm{strong}} \nonumber\\
  &=& e\,(1+1-2)\,V_{\mathrm{strong}} = 0.
\end{eqnarray}
This fixes the relative signs of the three channels,
$\sigma_{d}:\sigma_{p}:\sigma_{\bar 3}=+:+:-$. Here $Z_{\bar 3}\equiv-Z_{3}=-2$
is the all-incoming charge of the $\Hb$ leg. The magnitude $2$ resides in the
Ward-fixed coupling $Z_{3}e$, so the longitudinal parts enter weighted by the
all-incoming charges $(+1,+1,-2)$. The
magnetic and quadrupole currents are separately transverse and drop from the
test. Because the strong vertex is momentum dependent, however, a residual
momentum-mismatch survives, $k_{\mu}M^{\mu}_{\mathrm{fact}}\neq0$, and must be
cancelled by a contact current. The mismatch is carried entirely by the
tensor structure. $\Ctwo$ multiplies the only explicit derivative in the
strong vertex (acting on the deuteron field, Table~\ref{tab:couplings}).
The derivative-free $\Cone$ coupling is instead gauge invariant by the charge
cancellation alone. The required object is the
minimal covariant derivative ($\partial\to\partial-ieQA$) completion of the strong
vertex.
Its coefficient is fixed entirely by the deuteron charge and $\Ctwo$, with
no free parameter, and by construction
$k_{\mu}(M^{\mu}_{\mathrm{fact}}+\mathcal{B}_{W}^{\mu})=0$.

Contracted with the photon polarization, this boundary term, which we denote $\mathcal{B}_{W}$ for ``Ward'', has a closed
on-shell form in the same bracket variables as the rest of the amplitude.
With $\zeta$ the photon reference spinor and the positive-helicity photon, it is
\begin{widetext}
\begin{equation}
  \mathcal{B}_{W}^{+1} = -\,\frac{e Z_{d}\,\Ctwo}{2\,m_{d}^{3}\,\ad{\zeta}{\gamma}}\,
  \mathrm{sym}_{(I_{1}I_{2})}\Big\{
  \sq{\bm{d}^{I_{2}}}{\gamma}\big(\ad{\bm{h}}{\bm{d}^{I_{1}}}\ad{\zeta}{\bm{p}}
  +\ad{\bm{h}}{\zeta}\ad{\bm{d}^{I_{1}}}{\bm{p}}\big)
  +\ad{\bm{d}^{I_{1}}}{\zeta}\big(\sq{\bm{h}}{\bm{d}^{I_{2}}}\sq{\gamma}{\bm{p}}
  +\sq{\bm{h}}{\gamma}\sq{\bm{d}^{I_{2}}}{\bm{p}}\big)\Big\},
  \label{eq:BWbracket}
\end{equation}
\end{widetext}
and the negative-helicity form given by its parity image
($\ad{\,}{}\!\leftrightarrow\!\sq{\,}{}$, $\zeta\to\tilde\zeta$). 
We verify that adding $\mathcal{B}_{W}$ renders the amplitude
$\zeta$-independent component by component (all twelve massive-leg spin
states, both photon helicities) and enforces
$k_{\mu}\mathcal{M}^{\mu}=0$.

\subsection{Boundary Two-Body Currents}
\label{app:bdy}

Beyond the Ward boundary term $\mathcal{B}_{W}$, there are
transverse contact structures [Fig.~\ref{fig:diagrams}(d)], genuine
electromagnetic two-body currents, which are free LECs. We enumerate them with the
helicity-category algorithm of Ref.~\cite{DeAngelis:2022qco}. 
The minimal operator dimension is $6$ (two fermions at $\tfrac{3}{2}$, the deuteron at $1$, the photon field strength at $2$), at which the eight non-empty helicity
categories yield ten bracket structures in five parity pairs. Parity
conservation identifies the partners (under $\langle\,\rangle\!\leftrightarrow\![\,]$),
leaving five independent parity-even currents. With legs
$(1,2,3,4)=(\Hb,d,p,\gamma)$ and bold spinors on the three massive legs, the full
set is
\begin{align}
  \mathcal{B}_{1} &= \ad{\bm{1}}{\bm{2}}\ad{\bm{2}}{4}\ad{\bm{3}}{4}, &
  \mathcal{B}_{2} &= \ad{\bm{1}}{4}\ad{\bm{2}}{\bm{3}}\ad{\bm{2}}{4}, \nonumber\\
  \mathcal{B}_{3} &= \ad{\bm{1}}{4}\ad{\bm{2}}{4}\sq{\bm{2}}{\bm{3}}, &
  \mathcal{B}_{4} &= \ad{\bm{2}}{4}\ad{\bm{3}}{4}\sq{\bm{1}}{\bm{2}}, \nonumber\\
  \mathcal{B}_{5} &= \ad{\bm{2}}{4}^{2}\sq{\bm{1}}{\bm{3}}, &&
  \label{eq:boundary}
\end{align}
with dimensionless coefficients $c_{1},\dots,c_{5}$, each entering the
amplitude as $(e/\Lambda^{3})\,c_{a}\,\mathcal{B}_{a}$
(Table~\ref{tab:couplings}) and each manifestly
transverse since it is built from the photon field strength.
The leading magnetic-dipole two-body current, dominant for $\dpHe$ near
threshold, sits among these five. One derivative insertion raises the dimension to $7$ (explicit factor
$e/\Lambda^{4}$), populating the parity-complementary helicity
categories and producing seven further parity-even currents.

The angle-integrated $S$-factor does not separately resolve all five of these
dimension-6 couplings. Reducing $|\mathcal{M}|^{2}$ into its channel-spin and
entrance-wave sectors, and integrating over the photon angle (under which the
multipoles add incoherently), collapses $c_{1}$--$c_{5}$ onto the surviving
photon multipoles. Up to overall normalization the projections are
\begin{align}
  c_{\Mone}^{(2)} &\propto -c_{1}+3c_{2}+3c_{3}-c_{4}+2c_{5}, \label{eq:cM1d}\\
  c_{\Mone}^{(4)} &\propto \phantom{-}c_{1}+c_{4}+c_{5}, \label{eq:cM1q}\\
  c_{\Eone}       &\propto -7c_{1}+9c_{2}-6c_{3}-2c_{4}-2c_{5}, \label{eq:cE1map}\\
  c_{M2}^{(a)}    &\propto \phantom{-}c_{1}\phantom{+9c_{2}-6c_{3}}-2c_{4}+3c_{5}, \label{eq:cM2a}\\
  c_{M2}^{(b)}    &\propto 4c_{1}\phantom{+9c_{2}-6c_{3}}-c_{4}-2c_{5}. \label{eq:cM2b}
\end{align}
The two magnetic-dipole projections (the $^{2}S_{1/2}$ doublet and $^{4}S_{3/2}$
quartet) are orthogonal and of equal threshold weight, and $\mathcal{B}_{2},
\mathcal{B}_{3}$ are inert in the magnetic quadrupole. The map is one-to-one: the
five contact terms fill exactly the five structures $\{\Mone\,\text{doublet},\,
\Mone\,\text{quartet},\,\Eone,\,M2^{(a)},\,M2^{(b)}\}$ (the projection
determinant is nonzero). There is no electric-quadrupole partner: the $E2$
projection vanishes identically at dimension $6$. An explicit photon-multipole projection one dimension up locates the $E2$ contact at dimension $7$. There the
seven parity-even one-derivative currents, all of the field-strength--derivative
form $\langle i4\rangle\langle j4\rangle\langle k|p_{X}|l]$ and its parity
image, are \emph{entirely} electric-quadrupole, differing only in entrance wave,
channel spin, and $J$. The angle-integrated $S$-factor is sensitive to a single
combination of them, the spin-independent doublet $c_{\Etwo}$ of
Eq.~\eqref{eq:Udefs}. The remaining six are higher entrance-wave and quartet
structures that contribute only to polarization observables. The dimension-6 combinations above are what enter the matrix element of
Appendix~\ref{app:me}. The orthogonal directions in coefficient space, the
$\Mone$ doublet/quartet split and the $M2^{(a)}/M2^{(b)}$ split, are
directions to which the angle-integrated $S$-factor is degenerate (it sees
only the incoherent sums). Like the six dimension-7 combinations beyond
$c_{\Etwo}$, they are resolvable only by angular or polarization observables.

\subsection{Matrix Element}
\label{app:me}

The full tree amplitude to leading boundary order is the sum of the
factorized channels, the Ward boundary term $\mathcal{B}_{W}$, and the transverse
boundary currents,
\begin{equation}
  \mathcal{M} = \mathcal{M}^{\mathrm{fact}}
  + \mathcal{B}_{W}
  + \frac{e}{\Lambda^{3}}\sum_{a} c_{a}\,\mathcal{B}_{a}.
\end{equation}
Squaring with
the initial spin average $\tfrac16$, summing over the deuteron polarizations, and integrating over the photon
emission angle, produces an incoherent sum of multipoles
\begin{equation}
\begin{split}
  \int d\Omega\,\Mbar = \frac{4\pi}{3}
  \Big[\,& \mathcal{A}_{\Mone} + \mathcal{A}_{\Etwo}^{D}
    + \big(\mathcal{A}_{\Eone}+\mathcal{A}_{M2}\big)\tfrac{\prel^{2}}{\mh^{2}} \\
  & + \mathcal{A}_{\Etwo}\big(\tfrac{\prel^{2}}{\mh^{2}}\big)^{2}
  \Big].
\end{split}
  \label{eq:Msq}
\end{equation}
The $s$-wave magnetic dipole is constant at threshold, the $p$-wave electric dipole
rises as $\prel^{2}$ (alongside the subleading magnetic quadrupole
$\mathcal{A}_{M2}$ at the same $\prel^{2}$ order), and the $d$-wave electric
quadrupole as $\prel^{4}$.  The constant $\mathcal{A}_{\Mone}$ term fixes $S(0)$,
joined there by the small constant-in-energy $D$-state quadrupole
$\mathcal{A}_{\Etwo}^{D}$ (the $^{4}S_{3/2}$ entrance, below). The strengths
$\mathcal{A}_{\Eone}$ and $\mathcal{A}_{\Etwo}$ vanish as $\prel\to0$. Writing
each strength explicitly as the photon phase-space weight times the squared
reduced matrix element,
\begin{align}
  \mathcal{A}_{\Mone} &= \tfrac14\,\Egam^{2}\,\big|U_{\Mone}^{(2)}\big|^{2}
                       + \tfrac29\,\Egam^{2}\,\big|U_{\Mone}^{(4)}\big|^{2}, \label{eq:AM1}\\
  \mathcal{A}_{\Eone} &= \tfrac13\,\Egam^{2}\,\big|U_{\Eone}\big|^{2}, \label{eq:AE1}\\
  \mathcal{A}_{M2}    &= \Egam^{4}\Big(N_{M2}^{(a)}\big|U_{M2}^{(a)}\big|^{2}
                       + N_{M2}^{(b)}\big|U_{M2}^{(b)}\big|^{2}\Big), \label{eq:AM2}\\
  \mathcal{A}_{\Etwo} &= \tfrac13\,\Egam^{4}\,\big|U_{\Etwo}\big|^{2}
                       + \tfrac{2}{21}\,\Egam^{4}\,\big|U_{\Etwo}^{(D)}\big|^{2}, \label{eq:AE2}\\
  \mathcal{A}_{\Etwo}^{D} &= \tfrac{1}{15}\,\Egam^{4}\,\big|U_{\Etwo}^{(D)}\big|^{2}, \label{eq:AE2D}
\end{align}
where the dipoles carry $\Egam^{2}$ and the quadrupoles $\Egam^{4}$, the
kinematic curvature of Sec.~\ref{sec:sfactor}. The
magnetic-dipole weights $\tfrac14$ and $\tfrac29$ are the $^{2}S_{1/2}$ doublet
and $^{4}S_{3/2}$ quartet. The magnetic-quadrupole pair has relative weight
$N_{M2}^{(b)}/N_{M2}^{(a)}\simeq0.30$. Their absolute normalization, unlike the
ratio, is fixed only by the cross-multipole match, which we never need: $M2$
carries $\Egam^{4}$ against the dipoles' $\Egam^{2}$, so at the same entrance
order it is $\Egam^{2}$-suppressed relative to $E1$, and the fit is sensitive
only to a single effective $M2$ strength (Sec.~\ref{sec:results}).
The reduced matrix elements are bilinear in the three-point couplings (each
factorized term pairs $\CSv$ or $\CDv$ with a photon-vertex coupling of
Table~\ref{tab:couplings}) and linear in the boundary multipole contact terms of
Eqs.~\eqref{eq:cM1d}--\eqref{eq:cM2b},
\begin{align}
  U_{\Mone}^{(2)} &= \frac{\CSv}{m_{d}}\big(2\mu_{3}P_{3}+\tfrac43 \mu_{d}P_{d}
                     -\tfrac23 \mu_{p}P_{p}\big)+\frac{e}{\Lambda^{3}}\,c_{\Mone}^{(2)}, \label{eq:UM1d}\\
  U_{\Mone}^{(4)} &= \frac{\CSv}{m_{d}}\big(\mu_{d}P_{d}-2\mu_{p}P_{p}\big)+\frac{e}{\Lambda^{3}}\,c_{\Mone}^{(4)}, \label{eq:UM1q}\\
  U_{\Eone}       &= \frac{\eeff\,\CSv}{m_{d}}+\frac{e}{\Lambda}\,c_{\Eone}, \label{eq:UE1app}\\
  U_{M2}^{(a)}    &= \frac{e}{\Lambda^{3}}\,c_{M2}^{(a)}, \qquad U_{M2}^{(b)} = \frac{e}{\Lambda^{3}}\,c_{M2}^{(b)}, \label{eq:UM2}\\
  U_{\Etwo}       &= \frac{q_{\mathrm{eff}}\,\CSv}{m_{d}}+\frac{e}{\Lambda}\,c_{\Etwo}, \label{eq:UE2app}\\
  U_{\Etwo}^{(D)} &= \frac{q_{\mathrm{eff}}\,\CDv}{m_{d}}. \label{eq:UE2D}
\end{align}
The magnetic dipole splits into the doublet $U_{\Mone}^{(2)}$ and quartet
$U_{\Mone}^{(4)}$, which add incoherently and each carry an independent short-range
contact (the quartet point-like piece is fed only by the $p$ and $d$ channels). The
magnetic quadrupole $U_{M2}^{(a,b)}$ are pure contact terms (the one-body impulse
$M2$ from the moment currents is power-suppressed and neglected here). Here
$P_{X}=1/(2m_{X}\Egam)$ are the threshold channel propagators
($P_{p}:P_{d}:P_{3}=2.993:1.497:1$), $\mu_{X}$ the total magnetic moments of
Eq.~\eqref{eq:totalmoment} (Table~\ref{tab:inputs}), and
$\eeff,q_{\mathrm{eff}}$ the recoil charges of Eqs.~\eqref{eq:eeff}
and~\eqref{eq:qeff}. Every point-like piece is fed by the $s$-wave coupling
$\CSv$ of Eq.~\eqref{eq:CSCD}. The bracket structures
$\mathcal{E}^{(1)},\mathcal{E}^{(2)}$ are both predominantly $^{2}S_{1/2}$, so it
is their sum $\CSv=\tfrac12(\Cone+\Ctwo)$ that
normalizes the magnetic and electric dipoles. The tensor coupling $\Ctwo$
thus enters the dipoles coherently through $\CSv$ rather than as a separate
incoherent admixture. The orthogonal combination $\CDv=\tfrac12(\Cone-\Ctwo)$ is the
$D$-state proper.
It contributes at two orders: a constant-in-energy threshold term $\mathcal{A}_{\Etwo}^{D}$ from
the $^{4}S_{3/2}$ entrance [Eq.~\eqref{eq:AE2D}, weight $\tfrac{1}{15}$] and a
$d$-wave admixture to $\mathcal{A}_{\Etwo}$ from the $^{4}D_{3/2}$ and
$^{4}D_{5/2}$ entrances
[Eq.~\eqref{eq:AE2}, weight $\tfrac{2}{21}=\tfrac{1}{15}+\tfrac{1}{35}$,
i.e.\ $\tfrac27(C_{D}/C_{S})^{2}$
of the $S$-state quadrupole]. Both are of relative order
$(C_{D}/C_{S})^{2}\approx2\times10^{-3}$, well below the EFT truncation error,
but we carry them for completeness. The constant piece is the only contribution
to $S(0)$ beyond the magnetic dipole.

Folding Eq.~\eqref{eq:Msq} through two-body phase space and the
Coulomb treatment of Sec.~\ref{sec:sfactor} returns the $S$-factor of
Eq.~\eqref{eq:SE}.

\section{The Effective Three-Coefficient Fit}
\label{app:efffit}

\begin{figure}[t]
  \centering
  \includegraphics[width=\columnwidth]{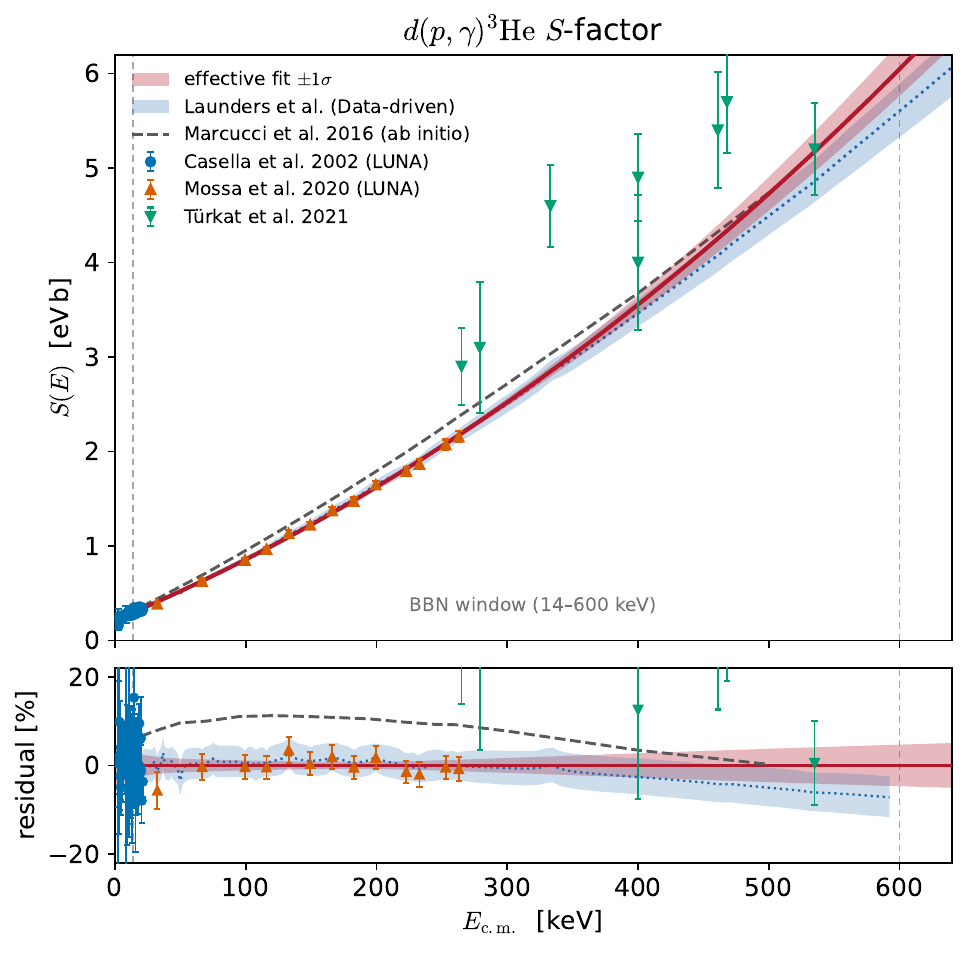}
  \caption{The $\dpHe$ astrophysical $S$-factor. The effective
three-coefficient fit (red band, $\pm1\sigma$) is shown with the
Casella~\cite{Casella:2002yej}, Mossa~\cite{Mossa:2020gjc}, and
T\"{u}rkat~\cite{Turkat:2021qmq} data. The {\it ab initio}
calculation~\cite{Marcucci:2015yla} (dashed) and the Gaussian-process band of
Launders {\it et al.}~\cite{Launders:2026ciu} (blue) are overlaid for
comparison. Lower panel: per-point residuals of the data relative to the fit.
  \label{fig:moneyplot}}
\end{figure}

This appendix reports the purely data-driven analysis drawn on in
Sec.~\ref{sec:results}: a fit of the three effective coefficients of
Eq.~\eqref{eq:poly}, independent of the theory inputs.

Once the fixed prefactor $P(E)$ of Eq.~\eqref{eq:poly} is divided out, the
$S$-factor is linear in $\{a_{0},a_{1},a_{2}\}$, so the fit is an ordinary
weighted linear least squares with a closed-form covariance. No minimization is
involved, and no input enters from the multipole normalization, the spin
multiplicities, or the asymptotic normalization. The three coefficients and
their covariance, together with the fixed kinematic $P(E)$, specify $S(E)$ over
the window---equivalently the conventional $S(0),S'(0),S''(0)$ of
Eq.~\eqref{eq:Sderiv}.

We fit the combined LUNA and T\"{u}rkat data over $E\le600\,\mathrm{keV}$, $73$
points in all. The Casella measurements~\cite{Casella:2002yej} reach the solar
Gamow peak and anchor the threshold $S(0)$; the Mossa
points~\cite{Mossa:2020gjc} span $32$--$263\,\mathrm{keV}$ and fix the
electric-dipole rise; the T\"{u}rkat data~\cite{Turkat:2021qmq} carry the
high-energy end. The documented overall scale uncertainties that lie outside the
per-point errors ($12\%$ for T\"{u}rkat, $\sim\!5\%$ for Casella) are treated as
correlated normalizations, one for each of these experiments. The
generalized-least-squares covariance adds $\delta_{e}^{2}\hat S_{i}\hat S_{j}$
within each experiment's block, with the correlated term evaluated on the model
prediction $\hat S$ to avoid the normalization bias of correlated
fits~\cite{DAgostini:1994zf}. The smaller Mossa
systematic is retained in its per-point error.

The fit gives $\chi^{2}/\mathrm{dof}=0.76$ over the $73$ points, with an
extrapolated threshold value
\begin{equation}
  S(0) = 0.214\pm0.010\,\mathrm{eV\,b},
  \label{eq:S0}
\end{equation}
in agreement with the global EFT fit of Sec.~\ref{sec:results}
($0.209\pm0.008\,\mathrm{eV\,b}$) and with the Solar Fusion~III recommendation
($0.203\pm0.005\,\mathrm{eV\,b}$~\cite{Acharya:2024lke}). With the T\"{u}rkat scale
carried as a correlated normalization, the precise LUNA data set the scale of
the curve. The fit then prefers a T\"{u}rkat normalization roughly twenty
percent below its central value. That shift is larger than the quoted $12\%$
scale---a genuine tension between the high-energy data and the LUNA-anchored
fit. The Gaussian-process posterior of Ref.~\cite{Launders:2026ciu} likewise
lies below the T\"{u}rkat points. Re-expressing the fit in the physics basis
that makes the photon weight $\Egam^{2}$ and the centrifugal barrier explicit
confirms the expectation of Sec.~\ref{sec:sfactor}: the curvature is dominated
by the exact kinematics. What remains of $S''(0)$ for the electric quadrupole
and the effective range is only about a percent.

Figure~\ref{fig:moneyplot} shows the best-fit $S(E)$ and its $\pm1\sigma$
band. The BBN window ($14$--$600\,\mathrm{keV}$) is marked by the dashed
vertical lines.

\bibliographystyle{apsrev4-2}
\bibliography{refs}

\end{document}